\documentclass[10pt,a4paper,notitlepage]{article}
\usepackage[activeacute,english]{babel}
\usepackage[latin1]{inputenc}
\usepackage[active]{srcltx}
\usepackage[toc,page]{appendix}
\usepackage{fancyhdr}
\usepackage{amsmath}
\usepackage{amsfonts}
\usepackage{amssymb}
\usepackage[mathcal]{euscript}
\usepackage{graphicx}% Include figure files
\usepackage{caption}
\usepackage{subcaption}
\usepackage{xcolor}
\usepackage[normalem]{ulem}
\usepackage{authblk}

\usepackage{booktabs}

\makeatletter
\let\old@startsection=\@startsection
\let\oldl@section=\l@section
\renewcommand{\@startsection}[6]{\old@startsection{#1}{#2}{#3}{#4}{#5}{#6\mathversion{bold}}}
\renewcommand{\l@section}[2]{\oldl@section{\mathversion{bold}#1}{#2}}
\makeatother

\makeatletter \@addtoreset{equation}{section} \makeatother

\title{\textbf{Magnetic Phenomena in Holographic Superconductivity\\
with Lifshitz Scaling}}

\author{Aldo Dector\thanks{Email address: aldo.dector@gmail.com}}

\affil{\textit{ Departament d'Estructura i Constituents de la Materia}\\
\textit{Facultat de F\'isica, Universitat de Barcelona}\\
\textit{Av. Diagonal 647, 08028 Barcelona, Spain}}

\date{}

\begin{document}
\begin{titlepage}
    \maketitle

    \begin{abstract}
		
We investigate the effects of Lifshitz dynamical critical exponent $z$ on a family of minimal $D=4+1$ holographic superconducting models, with a particular focus on magnetic phenomena. We see that it is possible to have a consistent Ginzburg-Landau approach to holographic superconductivity in a Lifshitz background. By following this phenomenological approach we are able to compute a wide array of physical quantities. We also calculate the Ginzburg-Landau parameter for different condensates, and conclude that in systems with higher dynamical critical exponent, vortex formation is more strongly unfavored energetically and exhibit a stronger Type I behavior. Finally, following the perturbative approach proposed by Maeda, Natsuume and Okamura, we calculate the critical magnetic field of our models for different values of $z$.

    \end{abstract}
  \end{titlepage}

\clearpage
%%%%%%%%%%%%%%%%%%%%%%%%%%%%%%%%%%%%%%%%%%%%%%%%%%%%%%%

\tableofcontents

\newpage

%%%%%%%%%%%%%%%%%%%%%%%%%%%%%%%%%%%%%%%%%%%%%%%%%%%%%%%%

\section{Introduction}

The AdS/CFT correspondence \cite{Aharony:1999ti} is one of the most important developments in theoretical physics in recent years. Because of the very nature of the duality, it also provides a promising new way of studying gauge theories in the strongly-coupled regime, where the usual perturbative methods fail to apply. The gauge/gravity duality has been used to gain insight in a wide variety of physical systems, such as the quark-gluon plasma or in condensed matter theory. In this last category, the gauge/gravity duality has been fruitfully applied in the study of high-$T_{c}$ superconductivity, where the usual BCS model of Cooper-pair creation ceases to be valid due to the strong interactions between the system's components. Because of these non-trivial interactions, holographic methods are hoped to shed some light in the understanding of these systems. This area of study,  known as \textit{holographic superconductivity}, is currently an exciting and very active area of research. (for some reviews, see, e.g. \cite{Hartnoll:2009sz, Horowitz:2010gk, McGreevy:2009xe, Musso:2014efa, Cai:2015cya}).

In very general terms, the usual models of holographic superconductivity are built around a local gauge group symmetry breaking by one of the component fields in the gravity side, where the gravitational solution is an asymptotically AdS charged black hole. This symmetry breaking in the gravity side signals the beginning of a superconducting phase in the dual field theory. (See, e.g. \cite{Gubser:2008px, Hartnoll:2008kx}). It has been found, however, that in some condensed matter systems phase transitions are governed by \textit{Lifshitz-like fixed points}. These exhibit the particular anisotropic spacetime scaling symmetry
\begin{equation}
\label{aniscaling}
t \rightarrow \lambda^{z}t\,,\hspace{20pt} x\rightarrow \lambda x\,,
\end{equation}
where $z$ is the \textit{Lifshitz dynamical critical exponent} governing the degree of anisotropy. This anisotropy breaks Lorentz invariance and the systems are non-relativistic in nature. Therefore, in order to study such field theories holographically, the dual gravitational description has to be modified. Indeed, it was found in \cite{Kachru:2008yh} that these Lifshitz-like fixed points can be described by the gravitational dual
\begin{equation}
\label{kachru}
ds^{2}=L^{2}\left(-r^{2z}dt^{2}+r^{2}dr^{2}+r^{2}\sum_{i=1}^{d}d\vec{x}^{2}\right)\,,
\end{equation}
which, for $z=1$ reduces back to the usual $AdS_{d+2}$ metric, but for $z\neq 1$ satisfies the anisotropic scaling (\ref{aniscaling}). A black hole generalization of this metric was found in \cite{Pang:2009ad}.

The purpose of this communication is to explore various aspects of holographic superconductivity with Lifshitz-like fixed points, with a particular focus on magnetic phenomena. We do this by starting from a minimal bulk model and by studying various choices of condensates. More concretely, we want to investigate how the dynamical critical exponent $z$ affects our system with respect to its behavior in the isotropic $z=1$ case. Most of the existing research on the subject was realized in $D=4$. See, for instance \cite{Brynjolfsson:2009ct, Sin:2009wi, Bu:2012zzb, Cai:2009hn}. In \cite{Lu:2013tza}, the authors do make an interesting treatment of the $D=5$ case, but have their interest put mainly on studying different kinds of superconductors (s-wave, p-wave, soliton) and on the computation of condensation and conductivity. Regarding the study of magnetic effects in a Lifshitz background, we note in particular \cite{Zhao:2013pva, Lala:2014jca}. In the first reference the authors also treat the $D=5$ case, but using a different condensate as the ones we will propose, and with a focus on the applicability of the matching method.

Although we will initially consider our minimal model in general dimensions, we will focus our attention on the $D=5$ case. The reason for this choice of dimension is that, as noted in \cite{Dias:2013bwa, Dector:2013dia}, dimensionality plays an important role in the way external magnetic fields act in the dual superconducting system. The standard argument is that in a $2+1$ ($D=4$) dimensional  superconductor an external 3+1 dimensional magnetic field will always penetrate the material because the energy needed to expel the field scales as the volume, while the energy that the system gains from being in a superconducting state scales as the area. This results in the system being a Type II superconductor. In the case of a $3+1$ ($D=5$) dimensional system such as the one we study, both energies scale as the volume and one has therefore a direct competition that does not excludes the possibility of obtaining a Type I superconductor. Also, while high-$T_{c}$ samples are typically composed of 2-dimensional $\text{CuO}_{2}$ layers (\textit{cuprate superconductors}), it is important to examine the effect of thickness when the system is probed by external magnetic fields.

 In this respect, in this paper we see that it is possible to have a consistent Ginzburg-Landau phenomenological  approach to holographic superconductivity \cite{Dector:2013dia} in a Lifshitz background. We then apply this Ginzburg-Landau approach to compute, among other physical quantities, the Ginzburg-Landau parameter of the system, and to see how it is affected by the dynamical critical exponent $z$. We will also study the effect of an external magnetic field acting directly on the system, using the approach proposed in \cite{Maeda:2009vf}. In order to have a more complete study of the system's properties, we managed to study a wide array of condensation cases, always within the $D=5$ framework, so that the general tendencies in the behavior of physical quantities become more clear.

The article is organized as follows. In section 2 we describe our minimal model of $D=5$ holographic superconductivity with Lifshitz dynamical scaling, and present the different cases of condensates we will consider. In section 3, we study small fluctuations around the component fields of our model. From these fluctuations we are able to compute the penetration and correlation lengths of the superconductor for our different cases. In section 4 we show that the system can be consistently described in a phenomenological Ginzburg-Landau approach. We compute the Ginzburg-Landau parameter $\kappa$ for the different cases in our model. In section 5, we compute the critical magnetic field $B_{c}$ of the superconductor. Finally, in section 6 we summarize the main results and discuss some open problems.

%%%%%%%%%%%%%%%%%%%%%%%%%%%%%%%%%%%%%%%%%%%%%%%%%%%%%%%%

\section{Minimal Holographic Superconductor in $D=d+2$ Lifshitz Background}

\subsection{General Setup}

\hspace{5pt}As mentioned in the introduction, the \textsl{D=d+2} gravitational dual (\ref{kachru}) can be generalized to a black hole solution \cite{Pang:2009ad}
\begin{equation}
\label{Pang}
ds^{2}=L^{2}\left(-r^{2z}f(r)dt^{2}+\frac{dr^{2}}{r^{2}f(r)}+r^{2}\sum_{i=1}^{d}dx_{i}^{2}\right)\,,
\end{equation}
where
\begin{equation}
f(r)=1-\frac{r_{h}^{z+d}}{r^{z+d}}\,,
\end{equation}
and where $r_{h}$ is the horizon of the black hole. The Lifshitz dynamical critical exponent can take values $1\leq z \leq d$. The gravitational solution (\ref{Pang}) can be obtained from the action \cite{Taylor:2008tg} 
\begin{equation}
\label{taylor}
S=\frac{1}{16\pi\, G_{d+1}}\int d^{d+2}x \sqrt{g}\left(R+\Lambda -\frac{1}{2}\left(\partial \varphi\right)^{2}-\frac{1}{4}e^{\lambda \varphi}\mathcal{F}^{2}\right)\,,
\end{equation}
with extremal solution for the fields
\begin{eqnarray}
\mathcal{F}_{rt}= q e^{-\lambda \varphi}\,,\hspace{20pt} e^{\lambda \varphi}=r^{\lambda \sqrt{2(z-1)d}}\,,\hspace{20pt}\nonumber\\
\lambda^{2}=\frac{2d}{z-1}\,,\hspace{20pt} q^{2}=2 L^{2}(z-1)(z+d)\,,\nonumber\\
\label{soltaylor}
\Lambda =-\frac{(z+d-1)(z+d)}{2L^{2}}\,,\hspace{50pt}
\end{eqnarray}
For the remaining of the paper we will set $L^{2}=1$. We will also prefer to work with the coordinate $u=r_{h}/r$. This change of coordinates gives
\begin{equation}
\label{gravsol}
ds^{2}=-\frac{r_{h}^{2z}f(u)}{u^{2z}}dt^{2}+\frac{1}{u^{2}f(u)}du^{2}+\frac{r_{h}^{2}}{u^{2}}\sum_{i=1}^{d} dx_{i}^{2}\,,
\end{equation}
where
\begin{equation}
\label{black}
f(u)=1-u^{z+d}\,,
\end{equation}
and the Hawking temperature is
\begin{equation}
T_{H}=\frac{(z+d)}{4\pi}r_{h}^{z}\,.
\end{equation}

 It is therefore the actions (\ref{taylor}) that will provide us the gravitational Lifshitz background (\ref{gravsol})-(\ref{black}).  We will now construct our minimal phenomenological model of holographic superconductivity by adding to (\ref{taylor}) the action term
\begin{equation}
S_{m}=\int d^{d+2}x \sqrt{-g}\left(-\frac{1}{4}F^{2}-\left|D\Psi\right|^{2}-m^{2}\left|\Psi\right|^{2}\right)\,,
\end{equation}
where we have introduced a charged scalar field $\Psi$ and a $U(1)$ gauge field $A_{\mu}$, following \cite{Hartnoll:2008kx}, and where $F_{\mu\nu}=\partial_{\mu}A_{\nu}-\partial_{\nu}A_{\mu}$, and $D_{\mu}=\nabla_{\mu}-i A_{\mu}$. We will assume that there is neglible interaction with the gravitational background and therefore it remains fixed and given by the Lifshitz black hole solution (\ref{gravsol})-(\ref{black}). This lack of back reaction means we are effectively working in the probe limit (very large scalar field charge). As we will explain below, the scalar field mass will be set so as to get particular dimensions for the condensate under study. 

The general equations of motion for these fields are 
\begin{eqnarray}
\label{psieom1}
D^{2}\Psi&=&m^{2}\Psi\,,\\
\label{phieom1}
\nabla_{\mu}F^{\mu \nu}&=&J^{\nu}+\left|\Psi\right|^{2}A^{\nu}\,,
\end{eqnarray}
where
\begin{equation}
J_{\mu}=i \left(\Psi^{*}\nabla_{\mu}\Psi - \Psi \nabla_{\mu}\Psi^{*}\right)\,.
\end{equation}
We  propose the following ansatz for the component fields
\begin{equation}
\label{fieldsanzats}
\Psi(u) = \frac{1}{\sqrt{2}}\psi(u)\,,\hspace{20pt} A = \phi(u)dt\,,
\end{equation}
where $\psi(u)$ is a real function. Under this ansatz the equations of motion (\ref{psieom1})-(\ref{phieom1}) become
\begin{eqnarray}
\label{EOMpsi}
\psi''+\left(\frac{f'}{f}-\frac{d+z-1}{u}\right)\psi'-\frac{1}{u^{2}f}\left(m^{2}-\frac{u^{2z}\phi^{2}}{r_{h}^{2z}f}\right)\psi&=&0\,,\\
\label{EOMphi}
\phi''-\frac{d-z-1}{u}\phi'-\frac{\psi^{2}}{u^{2}f}\phi&=&0\,.
\end{eqnarray} 
This system of equations admits the no-hair solution $\psi(r)=0$. In this case the gauge field has solutions 
\begin{eqnarray}
\label{phi0sol}
\phi(u)&=&\mu-\rho\, \frac{u^{d-z}}{r_{h}^{d-z}}\,,\hspace{44pt}(z\neq d)\,,\\
\phi(u)&=&\mu-\rho\,\text{log}\left(\frac{\xi r_{h}}{u}\right)\,,\hspace{20pt}(z=d)\,,
\end{eqnarray}
where $\xi$ is a constant. This no-hair solution will correspond to the normal phase of the superconductor. The superconducting phase will be given by solutions with $\psi(u)\neq 0$. From the equations of motion (\ref{EOMpsi})-(\ref{EOMphi}) we see that the asymptotic $u\rightarrow 0$ behavior of the fields is
\begin{equation}
\label{assympt}
\psi(u) \approx \mathcal{O}_{-}\frac{u^{\Delta_{-}}}{r_{h}^{\Delta_{-}}}+\mathcal{O}_{+}\frac{u^{\Delta_{+}}}{r_{h}^{\Delta_{+}}}+\cdots\,,\\
\end{equation}
and
\begin{eqnarray}
\label{asymptphiI}
\phi(u) &\approx& \mu -\rho \frac{u^{d-z}}{r_{h}^{d-z}}+\cdots\,,\hspace{44pt}(z\neq d)\,,\\
\phi(u) &\approx& \mu -\rho\, \text{log}\left(\frac{\xi r_{h}}{u}\right)+\cdots\,,\hspace{20pt}(z=d)\,,
\end{eqnarray}
with
\begin{equation}
\label{Delta}
\Delta_{\pm}=\frac{1}{2}\left((z+d)\pm\sqrt{(z+d)^{2}+4 m^{2}}\right)\,,
\end{equation}
from where we get the BF-bound on the mass
\begin{equation}
m^{2}\geq-\frac{(z+d)^{2}}{4}\,.
\end{equation}

According to the AdS/CFT dictionary, the asymptotic coefficient $\mathcal{O}_{+}$ corresponds to an operator of dimension $\Delta_{+}$, while $\mathcal{O}_{-}$ corresponds to a source in the boundary theory. Meanwhile, $\mu$ and $\rho$ correspond to the chemical potential and the charge density of the dual field theory, respectively. In order to solve Eqs. (\ref{EOMpsi})-(\ref{EOMphi}) we will impose the regularity condition at the horizon $\phi(u=1)=0$. Also, from Eq. (\ref{EOMpsi}) we obtain the additional condition at $u=1$
\begin{equation}
\psi'(1)=\frac{m^{2}}{f'(1)}\psi(1)\,.
\end{equation}
Additionally, in order to simplify the numerical calculations, we will make use of the scaling symmetries
\begin{equation}
\label{scaleinvariance}
r \rightarrow a r\,, \;\;  t\rightarrow \frac{1}{a^{z}}t\,,\;\; x_{i} \rightarrow  \frac{1}{a}x_{i} \,,\;\; g \rightarrow a^{2} g\,,\;\; \phi \rightarrow a^{z} \phi\,.
\end{equation}

As explained in the Introduction, we will focus on the particular case $D=5$. This means we will take $d=3$. When looking for hairy solutions to the equations of motion one has two possible boundary conditions at $u\rightarrow 0$. One can set $\mathcal{O}_{-}=0$ (\textit{standard quantization condition}), in which case one has the operator $\mathcal{O}_{+}$ (dimension $\Delta_{+}$) as the superconductor order parameter. Conversely, one can set the boundary condition $\mathcal{O}_{+}=0$ (\textit{alternative quantization condition}), which results in the operator $\mathcal{O}_{-}$ (dimension $\Delta_{-}$) being the superconductor order parameter. Having set either one of these boundary conditions, one can proceed to solve the equations of motion through the shooting method.

\subsection{Different Cases of Condensation}

Going back to the allowed values for the dynamical critical exponent, we see that for the $D=5$ case we can have $1 \leq z\leq 3$. Throughout this paper, for both brevity and simplicity, we will choose to work with the integer values $z=1, 2$. This suits perfectly our primary objective, stated in the introduction, which is to have a general idea of how the dynamical critical exponent $z$ affects our holographic superconductor with respect to its behavior in the usual ($z=1$) isotropic realization of the gauge/gravity duality. As will be seen in the following, the general tendency in the behavior of the physical quantities of the system will be very clear when treating these values.

In order to have a more comprehensive  study of the effect of the dynamical critical exponent $z$ on our holographic superconductor, we will choose to work in the following cases:

\bigskip

$\bullet$ \textbf{Case I}. We set the value of the scalar field mass as
\begin{equation}
m^{2}=- 3\,z\,.
\end{equation}
 In this way, we have
\begin{equation}
 \Delta_{-}=z\,,\hspace{20pt}\Delta_{+}=3\,,
\end{equation}
so that the asymptotic behavior of the scalar field at $u\rightarrow 0$ is
\begin{equation}
\psi(u)\approx \mathcal{O}_{z}\frac{u^{z}}{r_{h}^{z}}+\mathcal{O}_{3}\frac{u^{3}}{r_{h}^{3}}+\cdots\,.
\end{equation}
In this case, we will set $\mathcal{O}_{z}=0$ for all values of $z$ considered, so that the superconducting order parameter of the system will be given by $\mathcal{O}_{3}$ of dimension 3.\footnote{The same condensate was used in \cite{Lu:2013tza}, but magnetic properties were not studied in that paper.}

\bigskip

$\bullet$ \textbf{Case II}. We set the scalar mass as
\begin{equation}
m^{2}=-(z+2)\,.
\end{equation}
This choice of mass results in
\begin{equation}
\Delta_{-}=1\,,\hspace{20pt}\Delta_{+}=z+2\,,
\end{equation} 
so that near $u\rightarrow 0$ we have
\begin{equation}
\psi(u)\approx \mathcal{O}_{1}\frac{u}{r_{h}}+\mathcal{O}_{z+2}\frac{u^{z+2}}{r_{h}^{z+2}}+\cdots.
\end{equation}
Here we will choose to set $\mathcal{O}_{z+2}=0$ and the order parameter of the superconductor will be given by $\mathcal{O}_{1}$ of dimension 1.

\bigskip

\begin{figure}
\centering
\textbf{Value of the condensate $\mathcal{O}_{\Delta}$ as a function of temperature, for different cases.}
\begin{subfigure}{.5\textwidth}
  \centering
 \begin{picture}(250,150)
\put(0,0){\includegraphics*[width=1\linewidth]{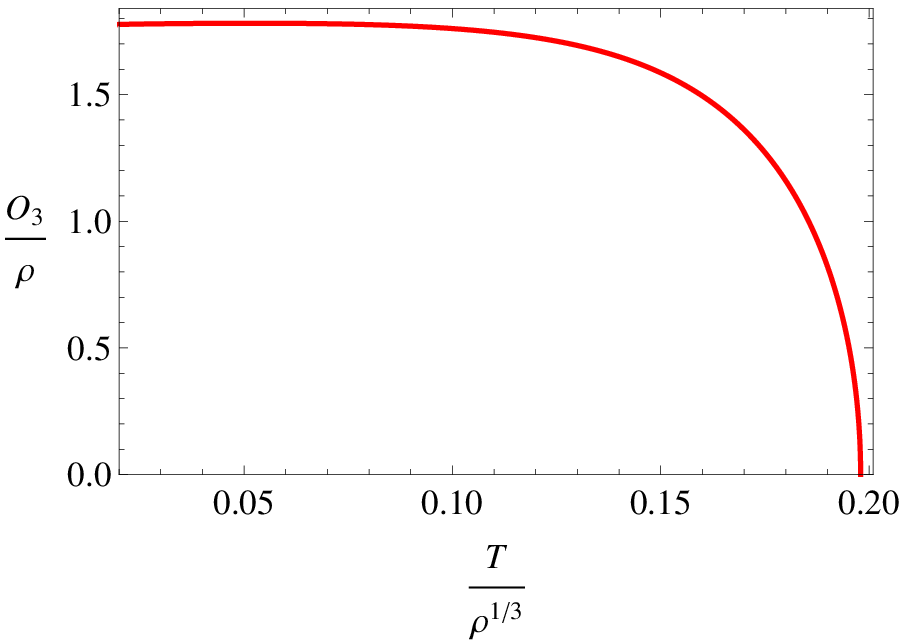}}%
\end{picture}
  \caption{$z=1$}
  \label{O3z1}
\end{subfigure}%
\begin{subfigure}{.5\textwidth}
  \centering
	\begin{picture}(250,150)
\put(10,0){\includegraphics*[width=1\linewidth]{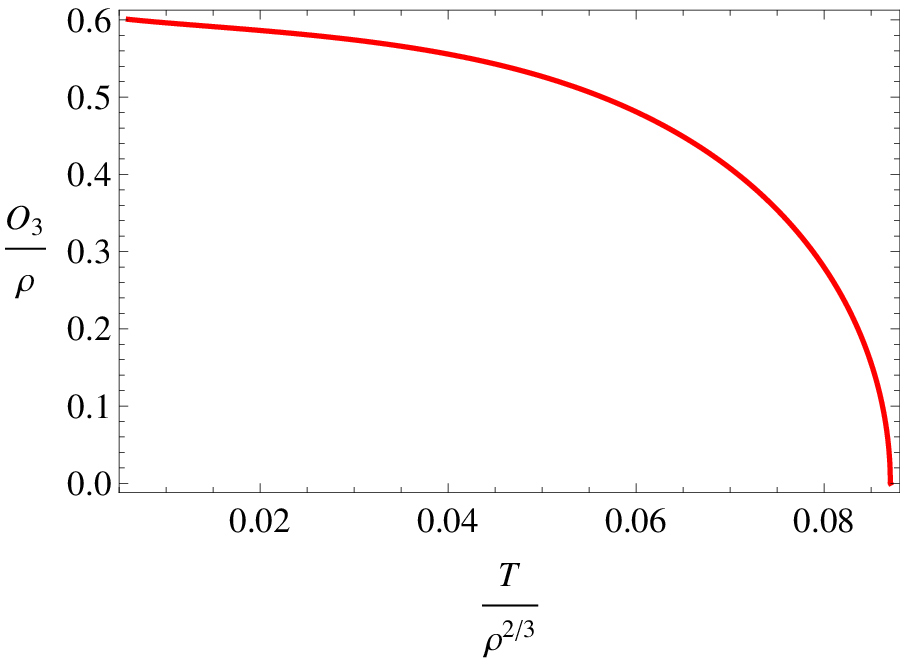}}%
\end{picture}
  \caption{$z=2$}
  \label{O3z2}
\end{subfigure}
\caption{\textbf{Case I}.}
\label{O3z}
\begin{subfigure}{.5\textwidth}
  \centering
 \begin{picture}(250,150)
\put(0,0){\includegraphics*[width=1\linewidth]{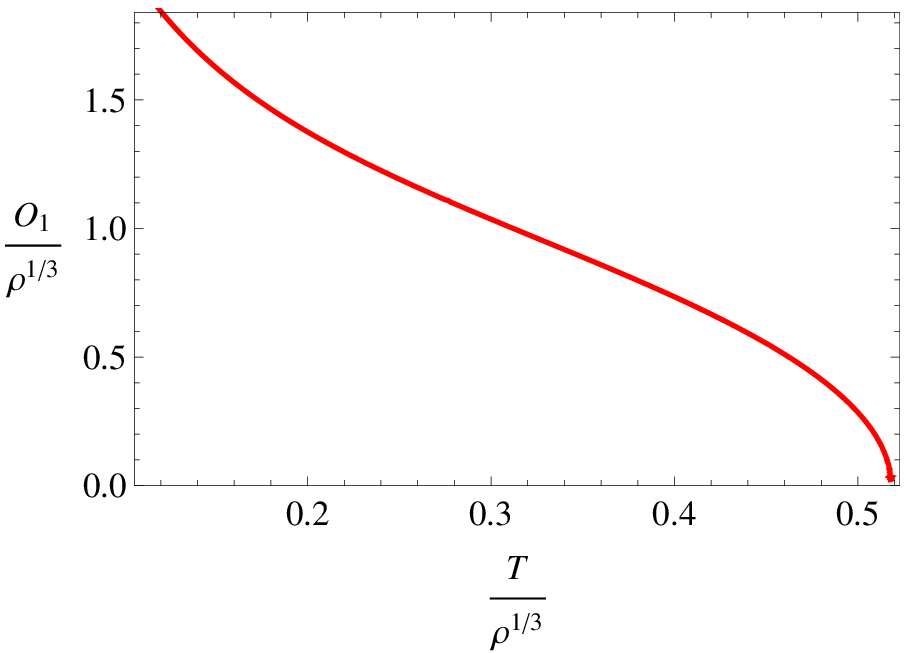}}%
\end{picture}
  \caption{$z=1$}
  \label{O1z1}
\end{subfigure}%
\begin{subfigure}{.5\textwidth}
  \centering
	\begin{picture}(250,150)
\put(10,0){\includegraphics*[width=1\linewidth]{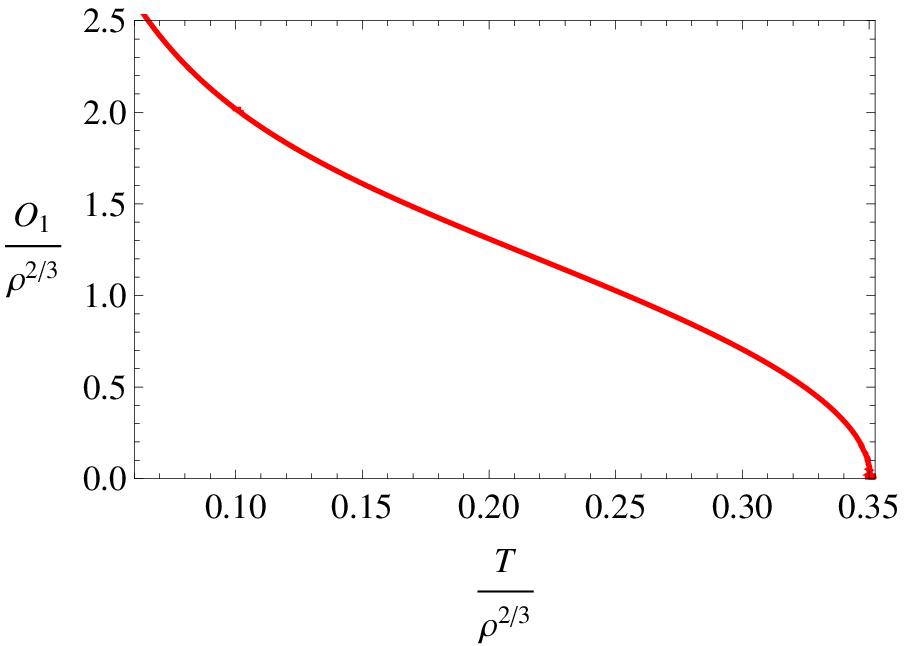}}%
\end{picture}
  \caption{$z=2$}
  \label{O1z2}
\end{subfigure}
\caption{\textbf{Case II}. }
\label{O1z}
\end{figure}

In figures \ref{O3z}-\ref{O1z} we show the value of the condensate $\mathcal{O}_{\Delta}$ as a function of temperature for each one of the cases described above. We notice that the near-$T_{c}$ the condensate behaves as
\begin{equation}
\mathcal{O}_{\Delta}\sim \left(1-T/T_{c}\right)^{1/2}\,,
\end{equation}
for all values of $z$. Therefore, the dynamical critical exponent does not alter the mean-field theory behavior of the order parameter. In Table \ref{tabTc} we show the value of the critical temperature $T_{c}$ for our different cases. We notice that the value of the critical temperature decreases with $z$ for all cases, and therefore a large dynamical critical exponent inhibits superconduction.

\begin{table}[t]
\begin{center}
\caption{\label{tabTc} Value of critical temperature $T_{c}$, for different cases. }
\begin{tabular}{ccc}
  \toprule
	\toprule
	$T_{c}/\rho^{z/3}$& $z=1$ & $z=2$ \\
	\toprule
  \textbf{Case I} & 0.198 & 0.087\\ 
  \textbf{Case II} & 0.517 & 0.351 \\
  \toprule
	\toprule
\end{tabular}
\end{center}
\end{table}

\section{Field Fluctuations}

\subsection{Gauge Field Fluctuation}

In this section we will add small fluctuations to the component fields of our model. As explained before, we will set $d=3$. We begin by adding the following gauge field fluctuation 
\begin{equation}
A = \phi(u)dt + \delta A_{x}(t,u,y)dx\,,
\end{equation}
where
\begin{equation}
\delta A_{x}(t,u,y)=e^{-i\omega t +i k y}A_{x}(u)\,.
\end{equation}
The corresponding equation of motion for $A_{x}(u)$ is, to linear order
\begin{equation}
\label{Axeq}
A_{x}''+\left(\frac{f'}{f}-\frac{d-z-3}{z}\right)A'_{x}+\left(\frac{u^{2z-2}\omega^{2}}{r_{h}^{2z}f^{2}}-\frac{k^{2}}{r_{h}^{2}f}-\frac{\psi^{2}}{z^{2}f}\right)A_{x}=0\,.
\end{equation}
We will consider the case where $\omega$ and $k$ are much smaller than the scale of the condensate (low-frequency/small-momentum regime), so that quadratic terms in $k$, $\omega$ in (\ref{Axeq}) can be neglected. Demanding regularity at the horizon $u=1$, from (\ref{Axeq}) we have the following conditions 
\begin{equation}
A_{x}(1)=A_{x0}\,,\hspace{20pt} A_{x}'(1)=\frac{\psi_{0}^{2}}{f'(1)}A_{x0}\,.
\end{equation}
where $\phi'(1)\equiv \psi_{0}$. Since Eq. (\ref{Axeq}) is linear, we will set $A_{x0}=1$ without loss of generality.

\begin{figure}
\centering
\textbf{Value of $\tilde{n}_{s}$ as a function of temperature, for different cases.}
\begin{subfigure}{.5\textwidth}
  \centering
 \begin{picture}(250,150)
\put(0,0){\includegraphics*[width=1\linewidth]{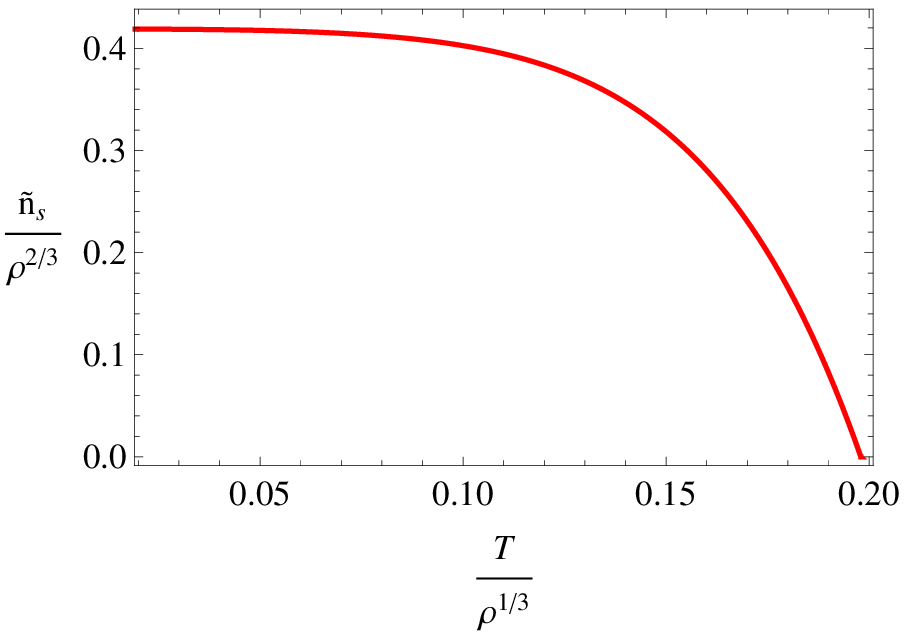}}%
\end{picture}
  \caption{$z=1$}
  \label{ns3z1}
\end{subfigure}%
\begin{subfigure}{.5\textwidth}
  \centering
	\begin{picture}(250,150)
\put(10,0){\includegraphics*[width=1\linewidth]{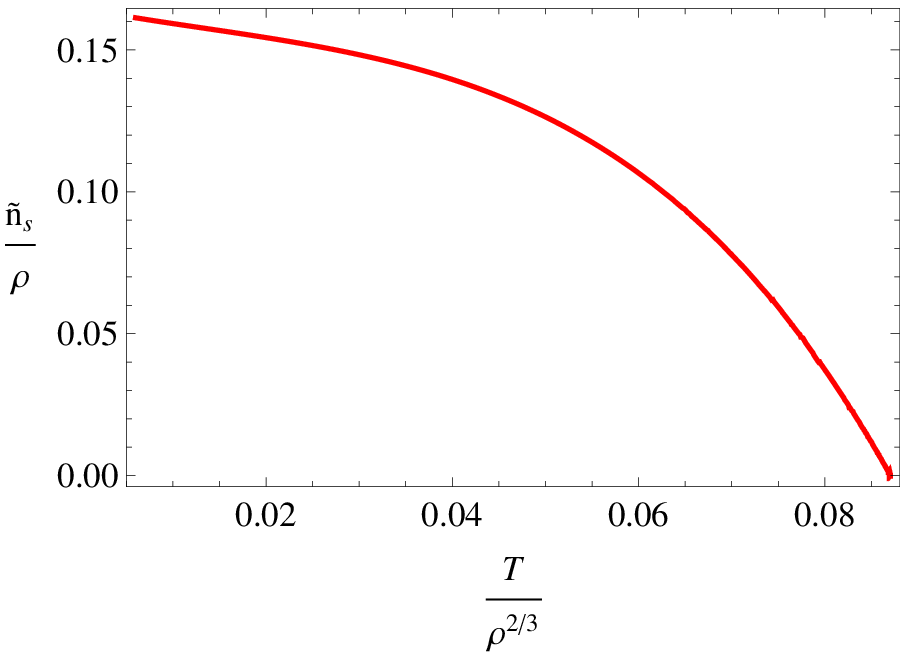}}%
\end{picture}
  \caption{$z=2$}
  \label{ns3z2}
\end{subfigure}
\caption{\textbf{Case I}. }
\label{ns3z}
\begin{subfigure}{.5\textwidth}
  \centering
 \begin{picture}(250,150)
\put(0,0){\includegraphics*[width=1\linewidth]{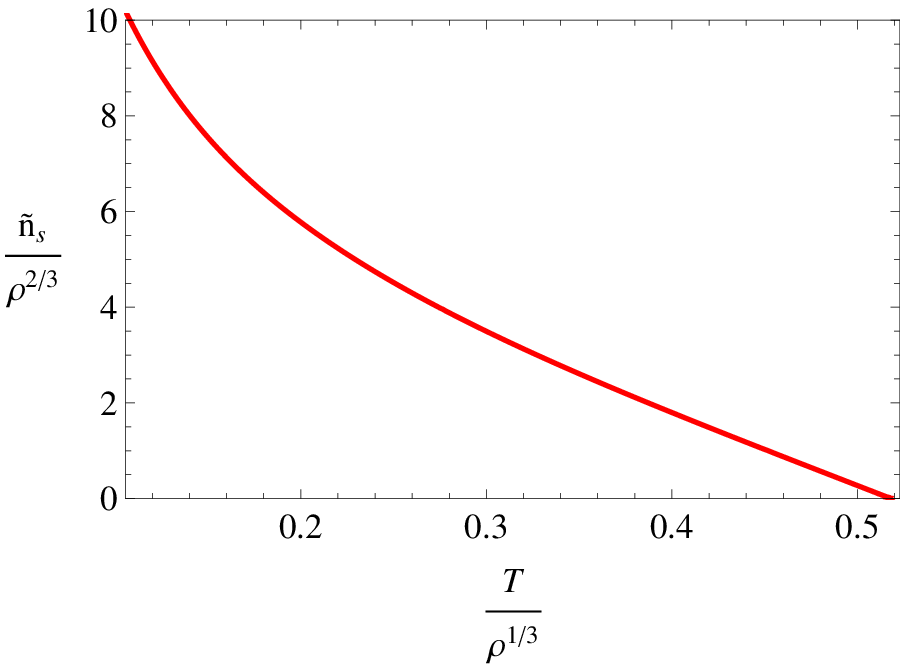}}%
\end{picture}
  \caption{$z=1$}
  \label{ns1z1}
\end{subfigure}%
\begin{subfigure}{.5\textwidth}
  \centering
	\begin{picture}(250,150)
\put(10,0){\includegraphics*[width=1\linewidth]{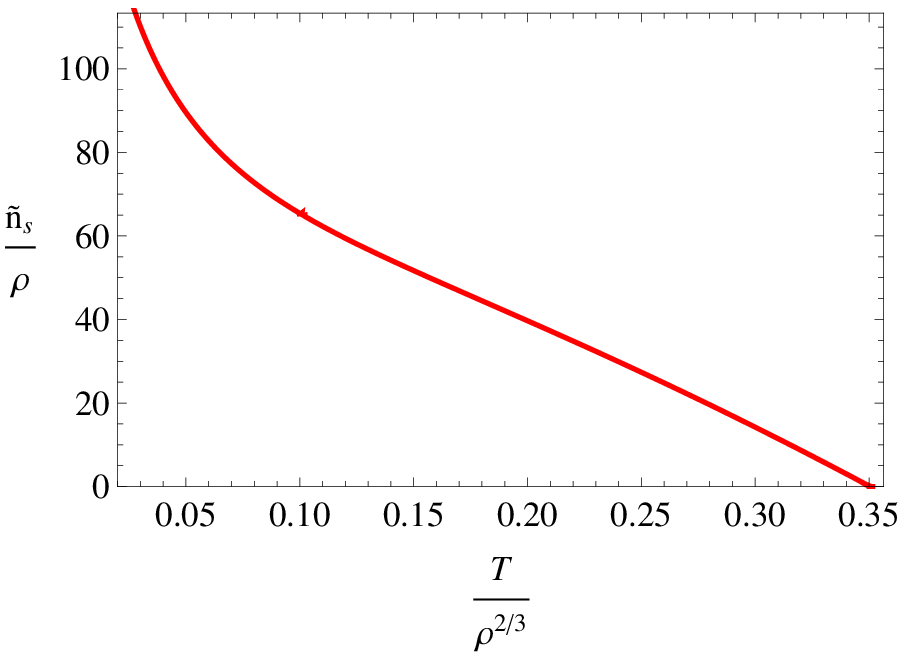}}%
\end{picture}
  \caption{$z=2$}
  \label{ns1z2}
\end{subfigure}
\caption{\textbf{Case II}.}
\label{ns1z}
\end{figure}

From (\ref{Axeq}) we see that the asymptotic behavior of $A_{x}$ at $u\rightarrow 0$ is
\begin{equation}
\label{asymptAx}
A_{x}(u)\approx A_{x}^{(0)}+J_{x}\frac{u^{d+z-2}}{r_{h}^{d+z-2}}+\cdots
\end{equation}
According to the AdS/CFT dictionary, $A_{x}^{(0)}$ corresponds to a vector potential in the dual field theory, while $J_{x}$ corresponds to its conjugate current \cite{Hartnoll:2008kx}. We can relate both physical quantities through the London equation 
\begin{equation}
\label{LondonC}
J_{x}=-\frac{1}{m}_{s}n_{s}A_{x}^{(0)}\,,
\end{equation}
where $n_{s}$ is the superconducting carrier density number and $m_{s}$ is the superconductor carrier mass. For simplicity we define the quantity
\begin{equation}
\label{tildens}
\tilde{n}_{s}\equiv \frac{1}{m_{s}}n_{s}\,,
\end{equation}
which can be computed holographically from (\ref{asymptAx}) and (\ref{LondonC}) as $ \tilde{n}_{s}=-J_{x}/A_{x}^{(0)}$. In figures \ref{ns3z}-\ref{ns1z} we show the value of $\tilde{n}_{s}$ as function of temperature, for different cases. We find that $\tilde{n}_{s}$ behaves near-$T_{c}$ as
\begin{equation}
\tilde{n}_{s}\sim \left(1-T/T_{c}\right)\,.
\end{equation}

\begin{figure}
\centering
\begin{subfigure}{.5\textwidth}
  \centering
 \begin{picture}(250,150)
\put(0,0){\includegraphics*[width=1.07\linewidth]{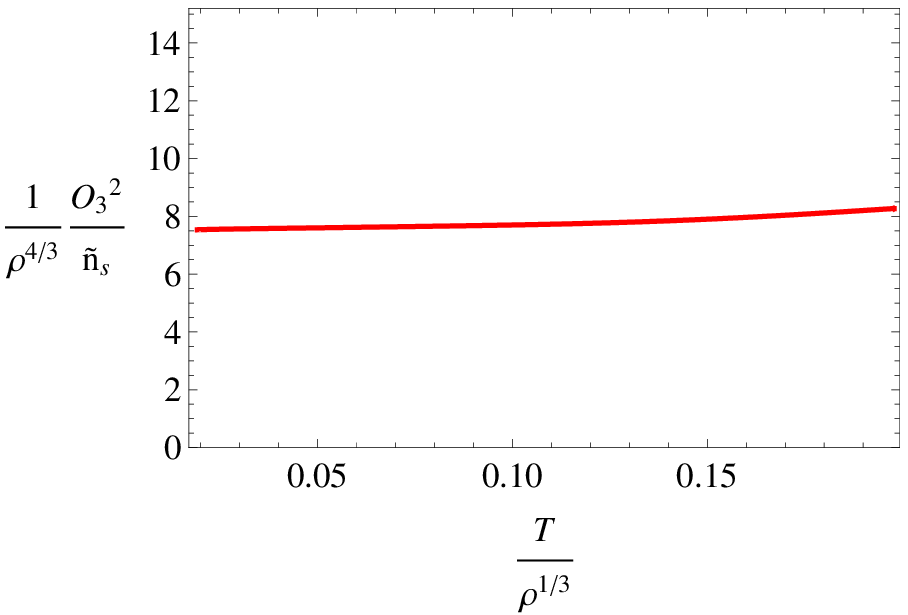}}%
\end{picture}
  \caption{\textbf{Case I}. $z=1$}
  \label{ratio3z1}
\end{subfigure}%
\begin{subfigure}{.5\textwidth}
  \centering
	\begin{picture}(250,150)
\put(17,2){\includegraphics*[width=1\linewidth]{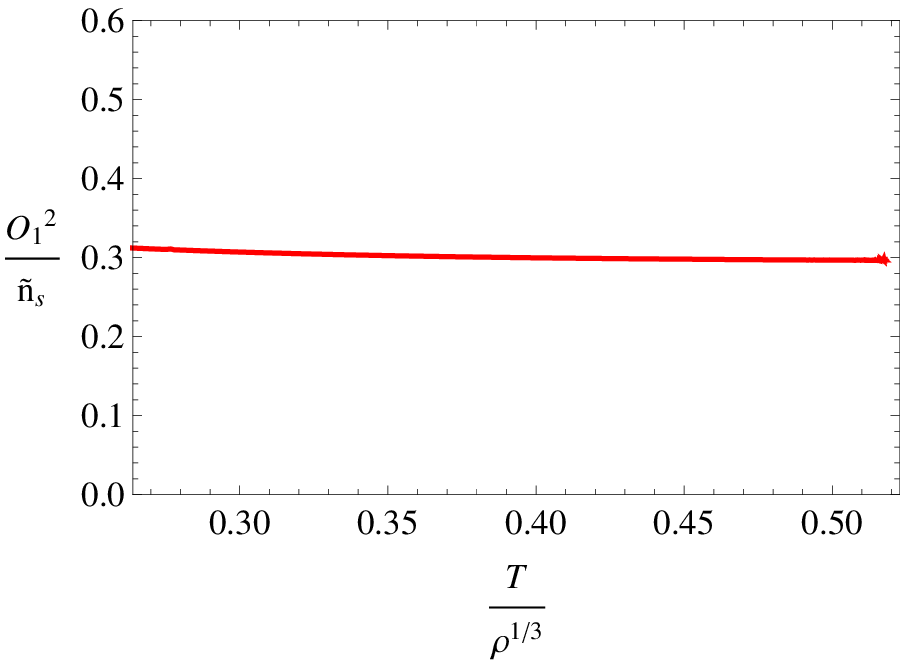}}%
\end{picture}
  \caption{\textbf{Case II}. $z=1$}
  \label{ratio1z1}
\end{subfigure}
\caption{Value of the ratio $\mathcal{O}_{\Delta}^{2}/\tilde{n}_{s}$ as a function of temperature, for different cases. }
\label{ratios}
\end{figure}

It is found numerically that the ratio of $\mathcal{O}_{\Delta}^{2}/\tilde{n}_{s}$ as a function of temperature behaves almost constantly and has a definite value at $T=T_{c}$ that varies according to the value of $z$ within a specific case of condensation. We define this ratio at $T_{c}$ as
\begin{equation}
\label{numericI}
\frac{\mathcal{O}_{\Delta}^{2}}{\tilde{n}_{s}}= C_{z}\,.
\end{equation}
We show in Table \ref{tabCz} how the constant $C_{z}$ varies for different cases. In figure \ref{ratios} we show this ratio as a function of temperature, for \textbf{Case I}, $z=1$, and \textbf{Case II}, $z=1$. We can observe that the ratio behaves almost like a constant with respect to $T$. The ratio (\ref{numericI}) will be important in the next section, when we apply the Ginzburg-Landau interpretation to our system.

\begin{table}[t]
\begin{center}
\caption{\label{tabCz} Value of $C_{z}$ for different cases.}
\begin{tabular}{ccc}
  \toprule
	\toprule
	$C_{z}/\rho^{(2\Delta-z-1)/3}$& $z=1$ & $z=2$\\
	\toprule
  \textbf{Case I} & 8.272 & 2.068\\ 
  \textbf{Case II} & 0.297 & 0.032\\
  \toprule
	\toprule
\end{tabular}
\end{center}
\end{table}

\subsection{Scalar Field Fluctuation}

We now consider a small fluctuation to the scalar field of the form
\begin{equation}
\Psi = \frac{1}{\sqrt{2}}\left(\psi(u) + \delta \psi(u,y)\right)\,, 
\end{equation}
with
\begin{equation}
\delta \psi(u,y)=e^{i k y}\eta(u)\,.
\end{equation}
The corresponding equation of motion is
\begin{equation}
\label{etaeq}
\eta''+\left(\frac{f'}{f}-\frac{(d+z-1)}{u}\right)\eta'-\frac{1}{u^{2}f}\left(m^{2}-\frac{u^{2z}\phi^{2}}{r_{h}^{2z}f}+\frac{u^{2}}{r_{h}^{2}}k^{2}\right)\eta=0\,,
\end{equation}
where we set $d=3$. Demanding regularity at the horizon $u=1$, from (\ref{etaeq}) we have the following conditions 
\begin{equation}
\eta(1)\equiv \eta_{0}\,,\hspace{20pt} \eta'(1)=\frac{1}{f'(1)}\left(m^{2}+\frac{k^{2}}{r_{h}^{2}}\right)\eta_{0}\,,
\end{equation}
while at $u\rightarrow 0$ we have the asymptotic behavior
\begin{equation}
\eta(u)\approx\left(\delta\mathcal{O}_{-}\right)\frac{u^{\Delta_{-}}}{r_{h}^{\Delta_{-}}}+\left(\delta\mathcal{O}_{+}\right)\frac{u^{\Delta_{+}}}{r_{h}^{\Delta_{+}}}+\cdots
\end{equation}
When solving equation (\ref{etaeq}) we set the same boundary conditions at $u\rightarrow 0$ as for the field $\psi$. Since we will not be concerned about the normalization of $\eta$, we set $\eta_{0}=1$.

\begin{figure}
\centering
\textbf{Value of the wave number $k$ as a function of temperature, for different cases.}
\begin{subfigure}{.5\textwidth}
  \centering
 \begin{picture}(250,150)
\put(0,0){\includegraphics*[width=1\linewidth]{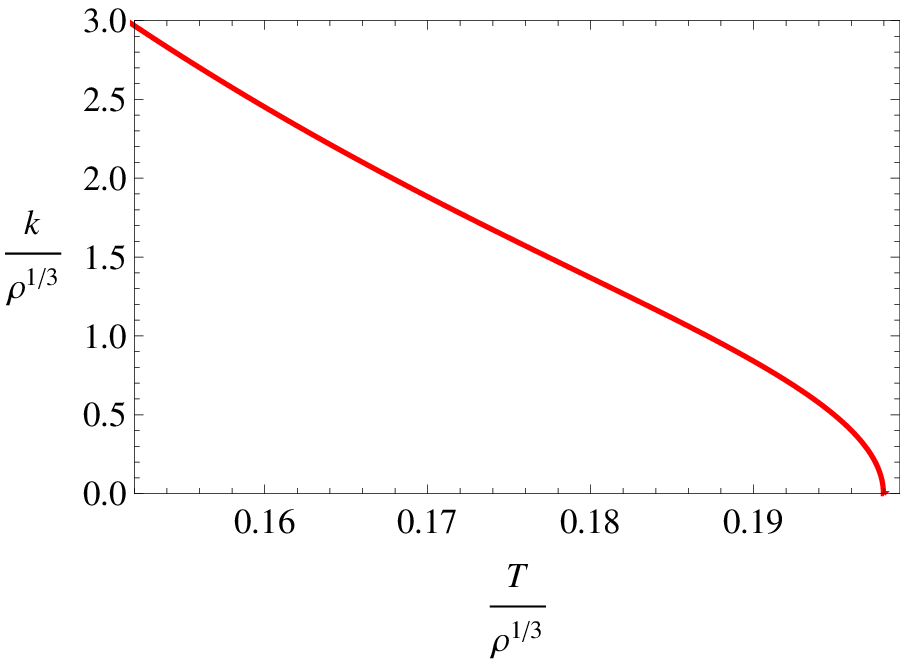}}%
\end{picture}
  \caption{$z=1$}
  \label{k3z1}
\end{subfigure}%
\begin{subfigure}{.5\textwidth}
  \centering
	\begin{picture}(250,150)
\put(10,0){\includegraphics*[width=1\linewidth]{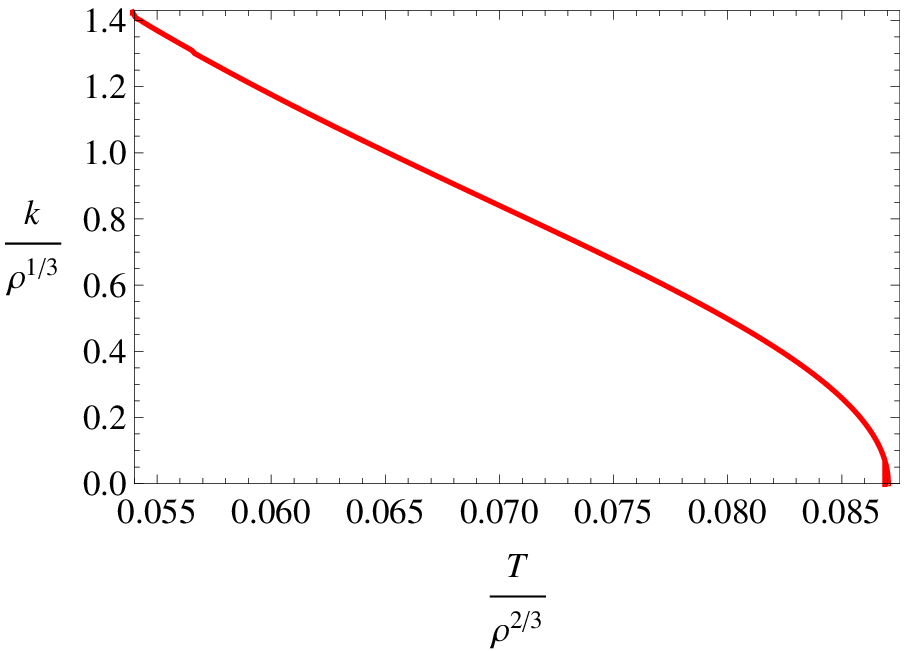}}%
\end{picture}
  \caption{$z=2$}
  \label{k3z2}
\end{subfigure}
\caption{\textbf{Case I}.}
\label{k3z}
\centering
\begin{subfigure}{.5\textwidth}
  \centering
 \begin{picture}(250,150)
\put(0,0){\includegraphics*[width=1\linewidth]{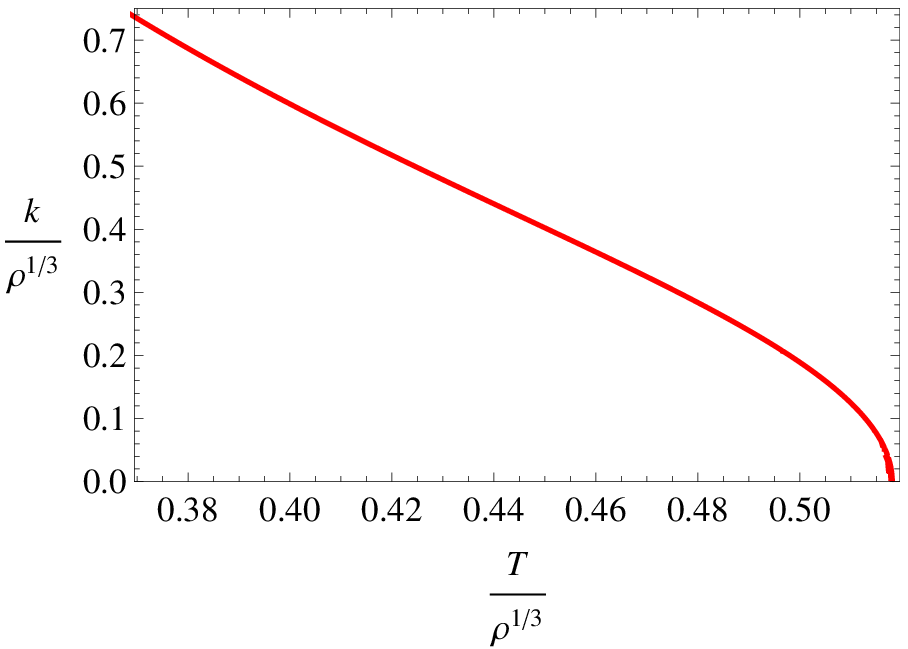}}%
\end{picture}
  \caption{$z=1$}
  \label{k1z1}
\end{subfigure}%
\begin{subfigure}{.5\textwidth}
  \centering
	\begin{picture}(250,150)
\put(10,0){\includegraphics*[width=1\linewidth]{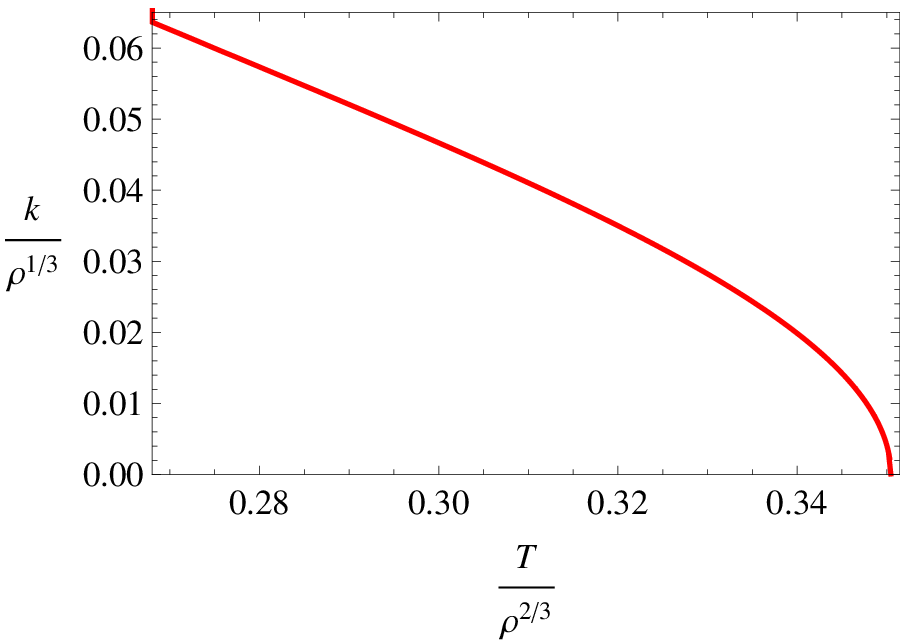}}%
\end{picture}
  \caption{$z=2$}
  \label{k1z2}
\end{subfigure}
\caption{\textbf{Case II}.}
\label{k1z}
\end{figure}

\begin{figure}
\centering
\textbf{Value of the correlation length $\xi_{0}$ as a function of temperature, for different cases.}
\begin{subfigure}{.5\textwidth}
  \centering
 \begin{picture}(250,150)
\put(0,0){\includegraphics*[width=1\linewidth]{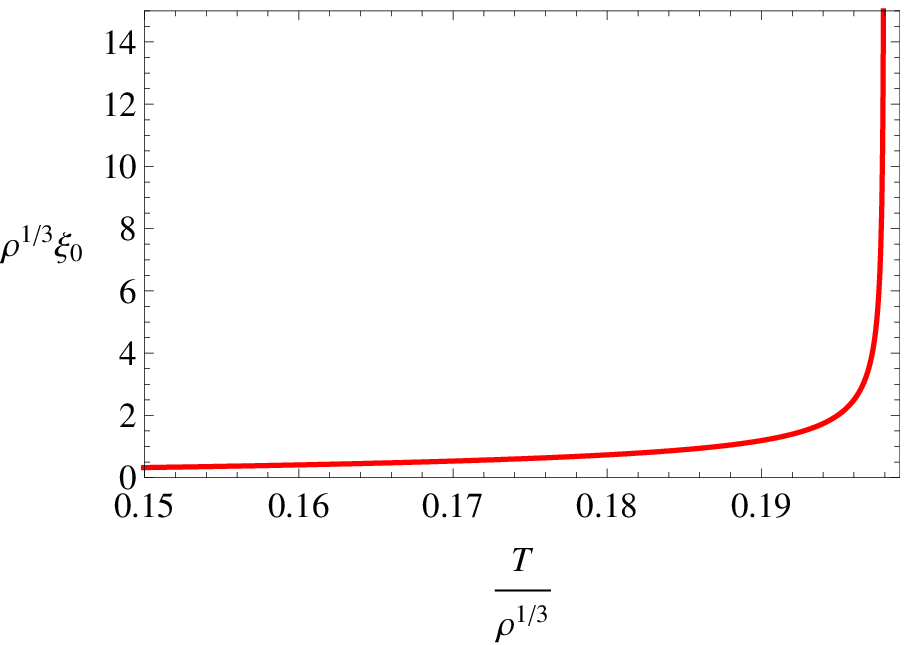}}%
\end{picture}
  \caption{$z=1$}
  \label{xi3z1}
\end{subfigure}%
\begin{subfigure}{.5\textwidth}
  \centering
	\begin{picture}(250,150)
\put(10,0){\includegraphics*[width=1\linewidth]{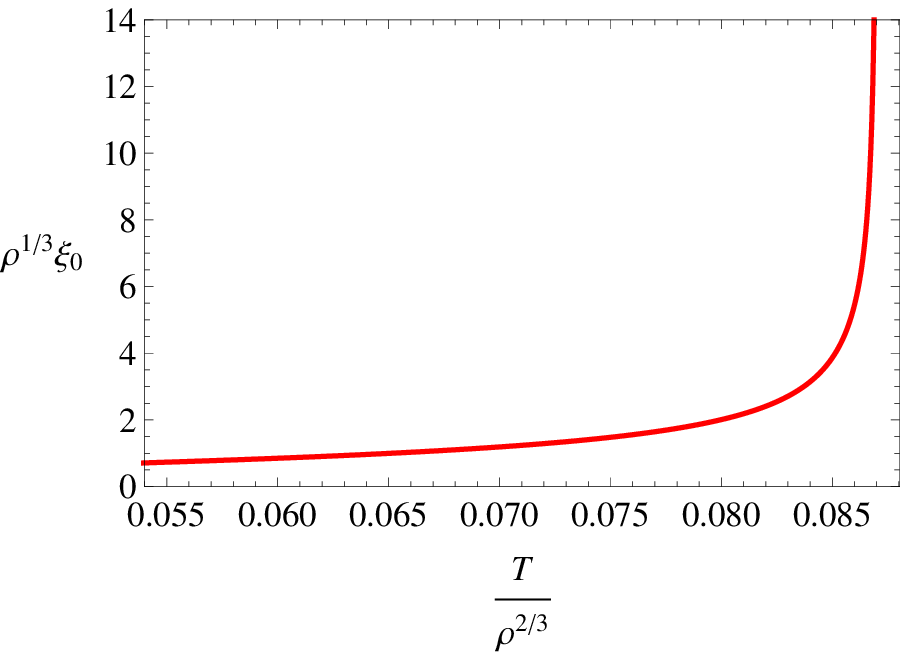}}%
\end{picture}
  \caption{$z=2$}
  \label{xi3z2}
\end{subfigure}
\caption{\textbf{Case I}.}
\label{xi3z}
\centering
\begin{subfigure}{.5\textwidth}
  \centering
 \begin{picture}(250,150)
\put(0,0){\includegraphics*[width=1\linewidth]{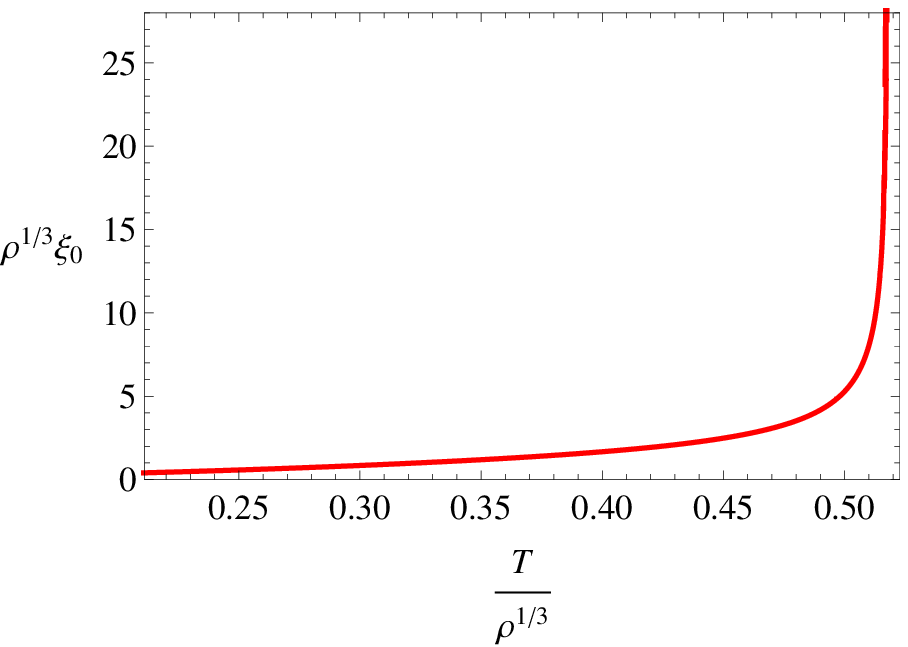}}%
\end{picture}
  \caption{$z=1$}
  \label{xi1z1}
\end{subfigure}%
\begin{subfigure}{.5\textwidth}
  \centering
	\begin{picture}(250,150)
\put(10,0){\includegraphics*[width=1\linewidth]{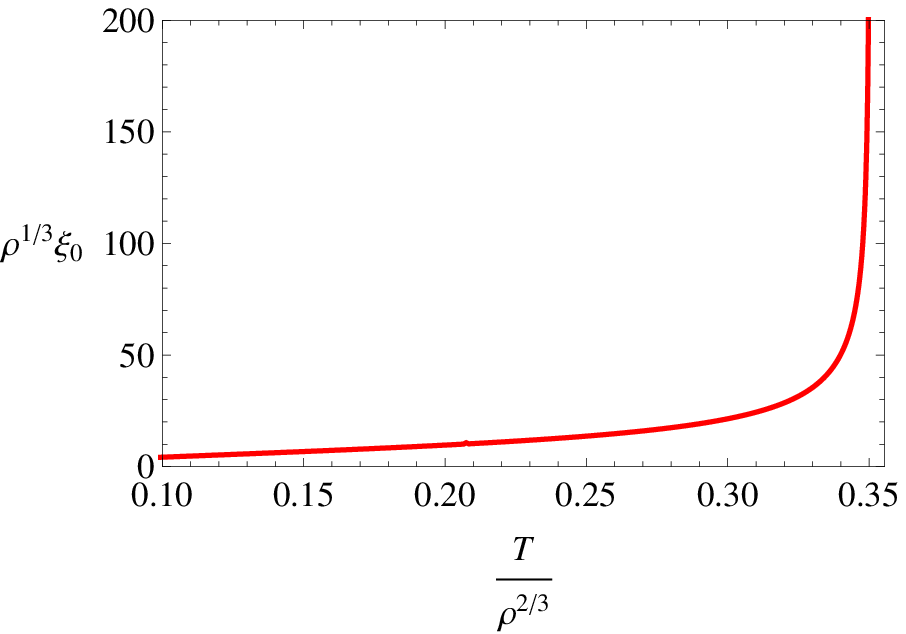}}%
\end{picture}
  \caption{$z=2$}
  \label{xi1z2}
\end{subfigure}
\caption{\textbf{Case II}.}
\label{xi1z}
\end{figure}

Following \cite{Maeda:2008ir}, we can compute holographically the correlation length of the boundary operator by calculating the wave number $k$. Indeed, the correlation length $\xi_{0}$ is the inverse of the pole of the correlation function of the order parameter written in Fourier space
\begin{equation}
\langle \mathcal{O}(k)\mathcal{O}(-k)\rangle \sim \frac{1}{|k|^{2}+1/\xi_{0}^{2}}\,.
\end{equation}
To obtain the wave number $k$, one must solve Eq. (\ref{etaeq}) as an eigenvalue problem consistent with the boundary conditions. This is done near the critical temperature. Once having computed $k$, one obtains the correlation length simply as
\begin{equation}
|\xi_{0}|=\frac{1}{|k|}\,.
\end{equation}
In figures \ref{k3z}-\ref{k1z} we show the value of $k$ as a value of temperature for our different cases. Also, in figures \ref{xi3z}-\ref{xi1z} we show the value of $\xi_{0}$ as a function of temperature, for our cases. We find that near the critical temperature, $k\sim \left(1-T/T_{c}\right)^{1/2}$, and equivalently
\begin{equation}
\xi_{0}\sim \frac{1}{\left(1-T/T_{c}\right)^{1/2}}\,,
\end{equation}
for all values of $z$, which is in agreement with mean field theory.

\section{Ginzburg-Landau Approach}

At this point we implement a phenomenological Ginzburg-Landau approach to our holographic superconductor, following \cite{Dector:2013dia}. The main assumption of this approach is that the dual field theory can be described near the critical temperature by the effective action
\begin{equation}
S_{\text{eff}}\approx \int dx^{4}\left\{\alpha \left|\Psi_{\text{GL}}\right|^{2}+\frac{\beta}{2}\left|\Psi_{\text{GL}}\right|^{4}+\frac{1}{2m_{s}}\left|D\Psi_{\text{GL}}\right|^{2}+\cdots\right\}\,.
\end{equation}
where the component fields are a scalar field $\Psi_{\text{GL}}$ representing the order parameter of the theory, and a vector field $\mathcal{A}_{\mu}$, with $\mu=0,...,3$ and where $D_{i}=\partial_{i}-i \mathcal{A}_{i}$. Also, $m_{s}$ is the superconductor carrier mass and $\alpha$, $\beta$ are phenomenological parameters with temperature dependence. According to the AdS/CFT dictionary, the vector field components $\mathcal{A}_{0}$ and $\mathcal{A}_{x}$ correspond to the chemical potential $\mu$ in (\ref{asymptphiI}) and to vector potential $A_{x}^{(0)}$ in (\ref{asymptAx}) respectively. According to mean field theory, the order parameter $\left|\Psi_{\text{GL}}\right|$ has critical exponent 1/2. In order to match this critical exponent with that of $\mathcal{O}_{\Delta}$ we propose the identification
\begin{equation}
\label{GLid}
\left|\Psi_{\text{GL}}\right|^{2}=N_{z}\mathcal{O}_{\Delta}^{2}\,,
\end{equation}
where $N_{z}$ is a proportionality constant that depends on $z$ and changes according to every model we consider.

In the remaining of this paper, we adopt the same notation and conventions of \cite{Dector:2013dia}. In particular, the superconducting carrier mass $m_{s}$ can be absorbed in definitions of the other parameters in Ginzburg-Landau theory, and we can therefore safely set $m_{s}=1$. Going back to (\ref{tildens}), this means in particular that $\tilde{n}_{s}=n_{s}$ and the numerical equality (\ref{numericI}) can be written as
\begin{equation}
\label{numericII}
\frac{\mathcal{O}_{\Delta}^{2}}{n_{s}}=C_{z}\,.
\end{equation}
However, one has according to Ginzburg-Landau theory the following relation between the order parameter $\left|\Psi_{\text{GL}}\right|$ and the charge carrier density $n_{s}$
\begin{equation}
\label{GLequ}
\left|\Psi_{\text{GL}}\right|^{2}=n_{s}\,.
\end{equation}
Then, substituting (\ref{GLequ}) and our identification (\ref{GLid}) in (\ref{numericII}) we obtain
\begin{equation}
N_{z}=\frac{1}{C_{z}}\,.
\end{equation}
In Table \ref{tabNz} we show the value of the proportionality constant $N_{z}$ for various cases.

\begin{figure}
\centering
\textbf{Value of the penetration length $\lambda$ as a function of temperature, for different cases.}
\begin{subfigure}{.5\textwidth}
  \centering
 \begin{picture}(250,150)
\put(0,0){\includegraphics*[width=1\linewidth]{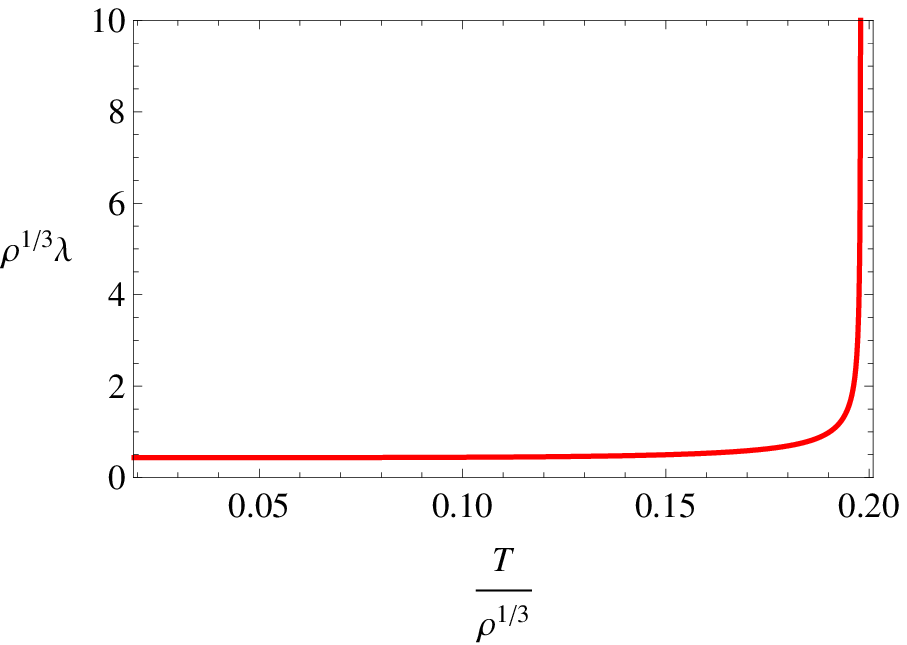}}%
\end{picture}
  \caption{$z=1$}
  \label{lambda3z1}
\end{subfigure}%
\begin{subfigure}{.5\textwidth}
  \centering
	\begin{picture}(250,150)
\put(10,0){\includegraphics*[width=1\linewidth]{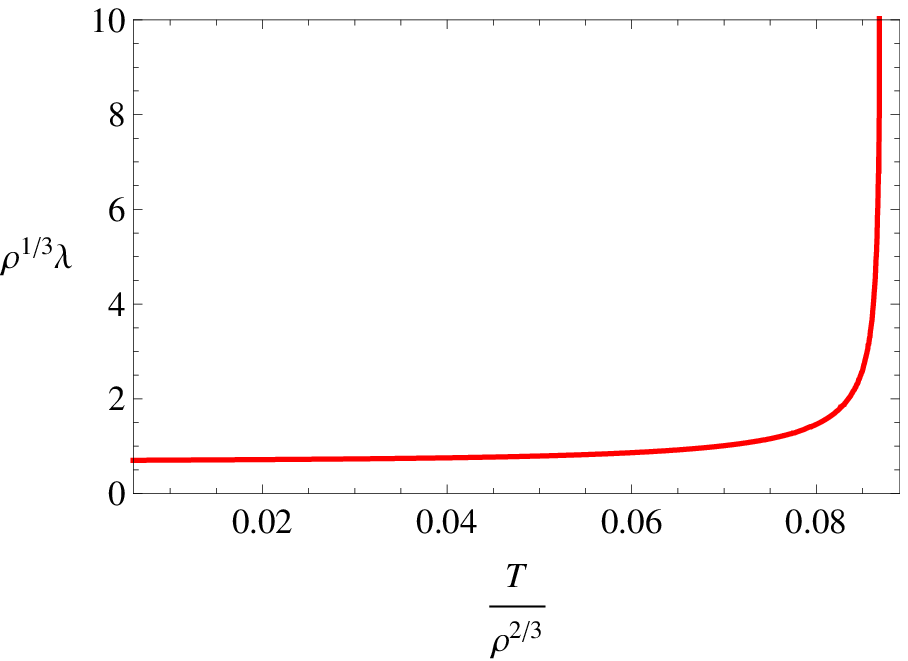}}%
\end{picture}
  \caption{$z=2$}
  \label{lambda3z2}
\end{subfigure}
\caption{\textbf{Case I}.}
\label{lambda3z}
\centering
\begin{subfigure}{.5\textwidth}
  \centering
 \begin{picture}(250,150)
\put(0,0){\includegraphics*[width=1\linewidth]{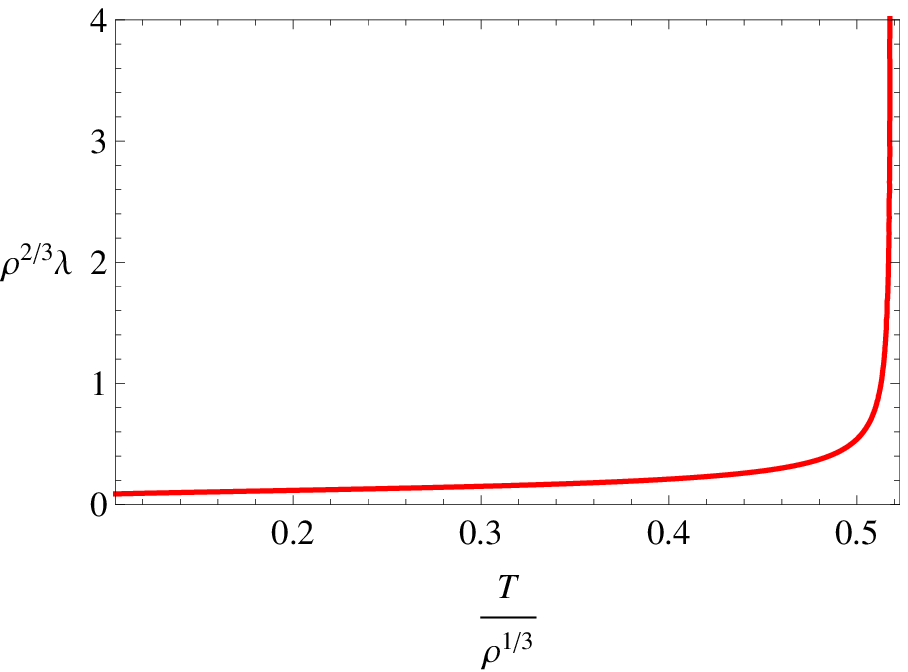}}%
\end{picture}
  \caption{$z=1$}
  \label{lambda1z1}
\end{subfigure}%
\begin{subfigure}{.5\textwidth}
  \centering
	\begin{picture}(250,150)
\put(10,0){\includegraphics*[width=1\linewidth]{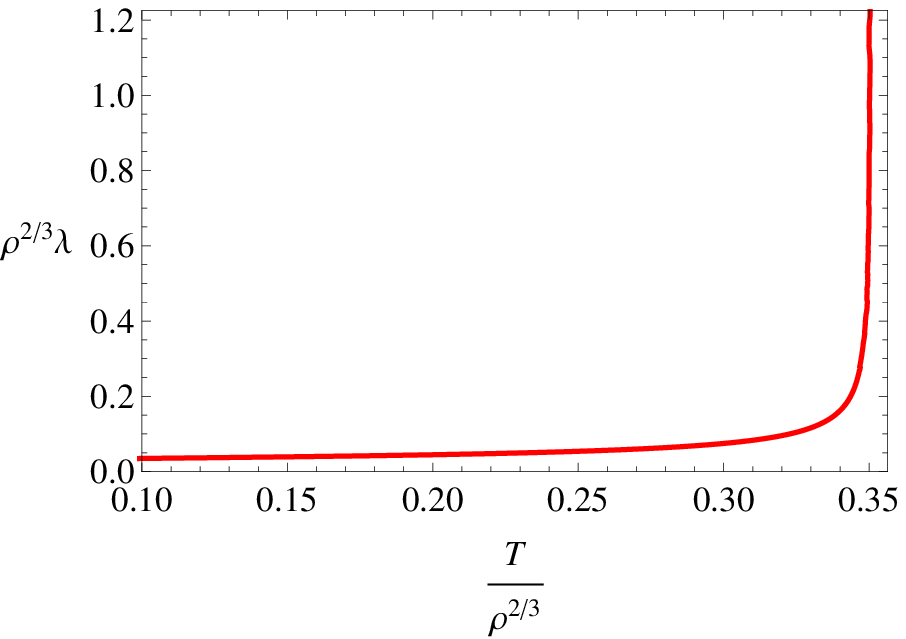}}%
\end{picture}
  \caption{$z=2$}
  \label{lambda1z2}
\end{subfigure}
\caption{\textbf{Case II}.}
\label{lambda1z}
\end{figure}

We can also calculate the penetration length $\lambda$ of the superconductor, defined as
\begin{equation}
\lambda = \frac{1}{\sqrt{4 \pi n_{s}}}\,.
\end{equation}
In figures \ref{lambda3z}-\ref{lambda1z} we show the value of $\lambda$ as a function of temperature, for our different cases. As in the case of $\xi_{0}$, we have that the behavior of $\lambda$ at $T\approx T_{c}$ is 
\begin{equation}
\lambda \sim \frac{1}{(1-T/T_{c})^{1/2}}\,,
\end{equation}
for all $z$. This result is in agreement with mean field theory.

\begin{table}[t]
\begin{center}
\caption{\label{tabNz} Value of $N_{z}$ for different cases.}
\begin{tabular}{ccc}
  \toprule
	\toprule
	$\rho^{(2\Delta-z-2)/3}N_{z}$& $z=1$ & $z=2$ \\
	\toprule
  \textbf{Case I} & 0.121 & 0.484\\ 
  \textbf{Case II} & 3.367 & 30.986\\
  \toprule
	\toprule
\end{tabular}
\end{center}
\end{table}

\begin{figure}
\centering
\textbf{Value of the parameter $\alpha$ as a function of temperature, for different cases.}
\begin{subfigure}{.5\textwidth}
  \centering
 \begin{picture}(250,150)
\put(0,0){\includegraphics*[width=1\linewidth]{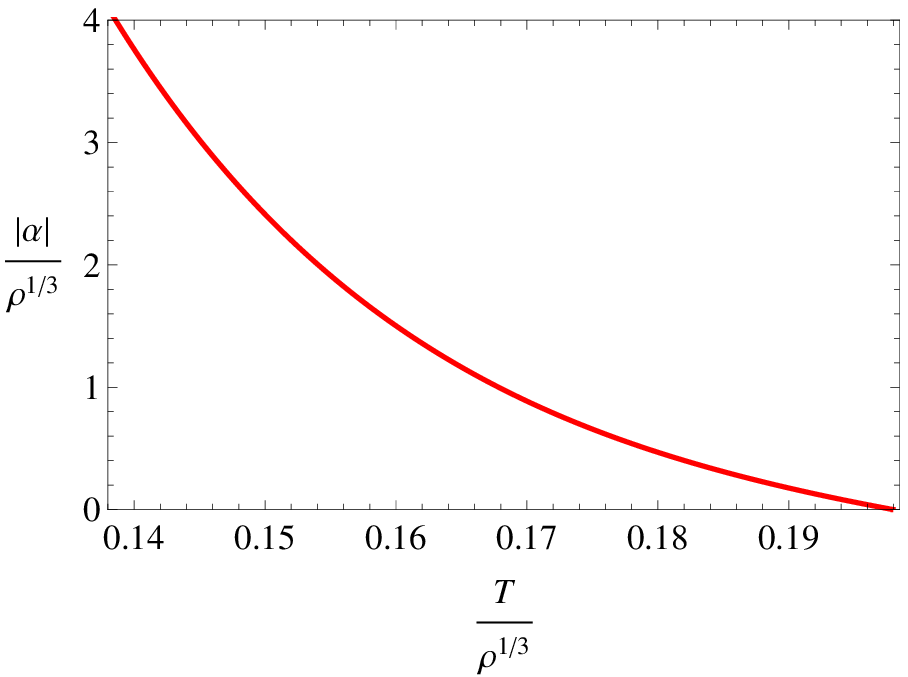}}%
\end{picture}
  \caption{$z=1$}
  \label{alpha3z1}
\end{subfigure}%
\begin{subfigure}{.5\textwidth}
  \centering
	\begin{picture}(250,150)
\put(10,0){\includegraphics*[width=1\linewidth]{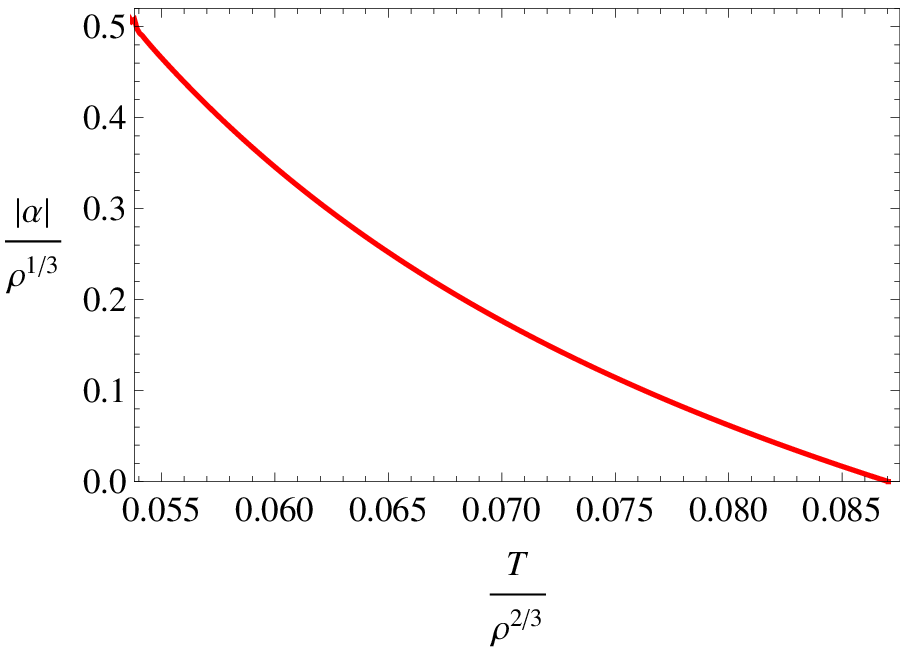}}%
\end{picture}
  \caption{$z=2$}
  \label{alpha3z2}
\end{subfigure}
\caption{\textbf{Case I}.}
\label{alpha3z}
\centering
\begin{subfigure}{.5\textwidth}
  \centering
 \begin{picture}(250,150)
\put(0,0){\includegraphics*[width=1\linewidth]{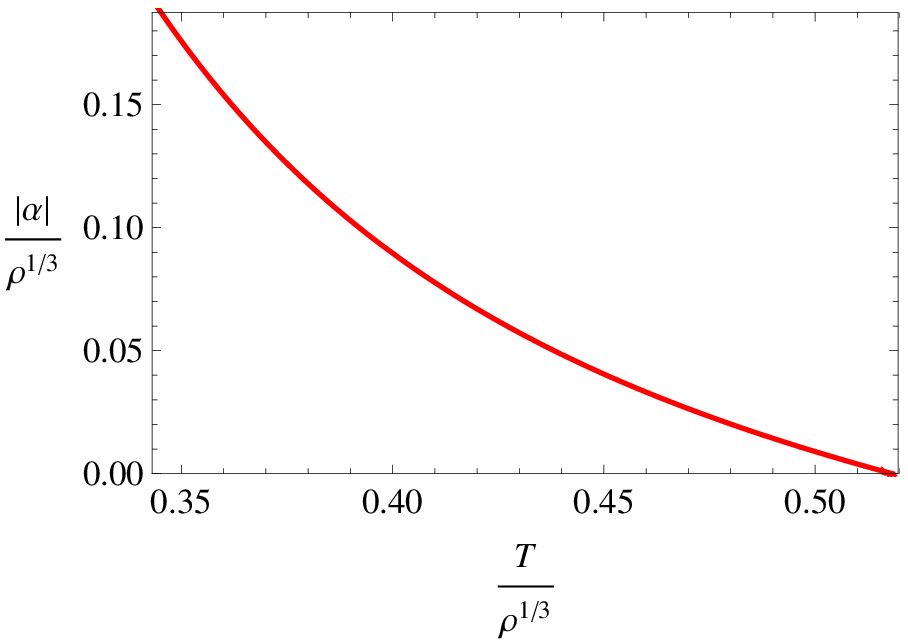}}%
\end{picture}
  \caption{$z=1$}
  \label{alpha1z1}
\end{subfigure}%
\begin{subfigure}{.5\textwidth}
  \centering
	\begin{picture}(250,150)
\put(10,0){\includegraphics*[width=1\linewidth]{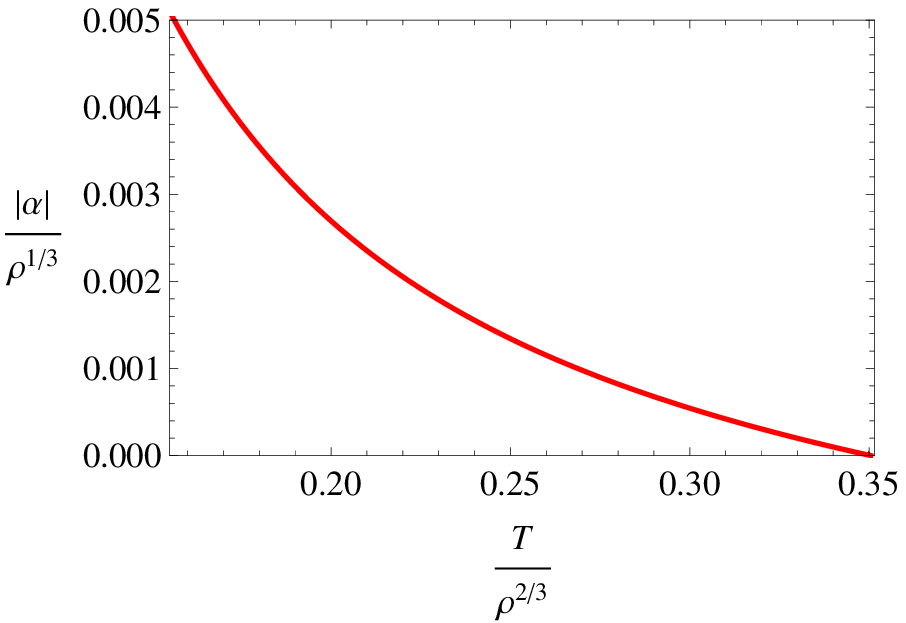}}%
\end{picture}
  \caption{$z=2$}
  \label{alpha1z2}
\end{subfigure}
\caption{\textbf{Case II}.}
\label{alpha1z}
\end{figure}

In order to have a consistent Ginzburg-Landau description of the dual field theory, we must be able to determine by holographic methods the parameters $\left|\alpha\right|$ and $\beta$. Regarding $\left|\alpha\right|$, we can determine it directly from the Ginzburg-Landau theory relation\footnote{The actual Ginzburg-Landau theory relation is
\begin{equation}
\left|\alpha\right|=\frac{\hbar^{2}}{2 m_{s}\xi^{2}}\,,
\end{equation}
where $\xi$ is the Ginzburg-Landau coherence length. The coherence length is in turn related to the correlation length $\xi_{0}$ as $\xi^{2}=2\xi_{0}^{2}$. See \cite{Dector:2013dia}.}
\begin{equation}
\label{alpha}
\left|\alpha\right|=\frac{1}{4\xi_{0}^{2}}\,.
\end{equation}
In figures \ref{alpha3z}-\ref{alpha1z} we show the value of $\left|\alpha\right|$ as a function of temperature, for our various cases. We see that the near-$T_{c}$ behavior of $\left|\alpha\right|$ is
\begin{equation}
\left|\alpha\right|\sim \alpha_{0}\left(1-T/T_{c}\right)\,,
\end{equation}
which is in agreement with usual Ginzburg-Landau theory, for all $z$. However, we find numerically that the  value of the coefficient $\alpha_{0}$ decreases as the value of $z$ raises.

\begin{figure}
\centering
\textbf{Value of the parameter $\beta$ as a function of temperature, for different cases.}
\begin{subfigure}{.5\textwidth}
  \centering
 \begin{picture}(250,150)
\put(0,0){\includegraphics*[width=1\linewidth]{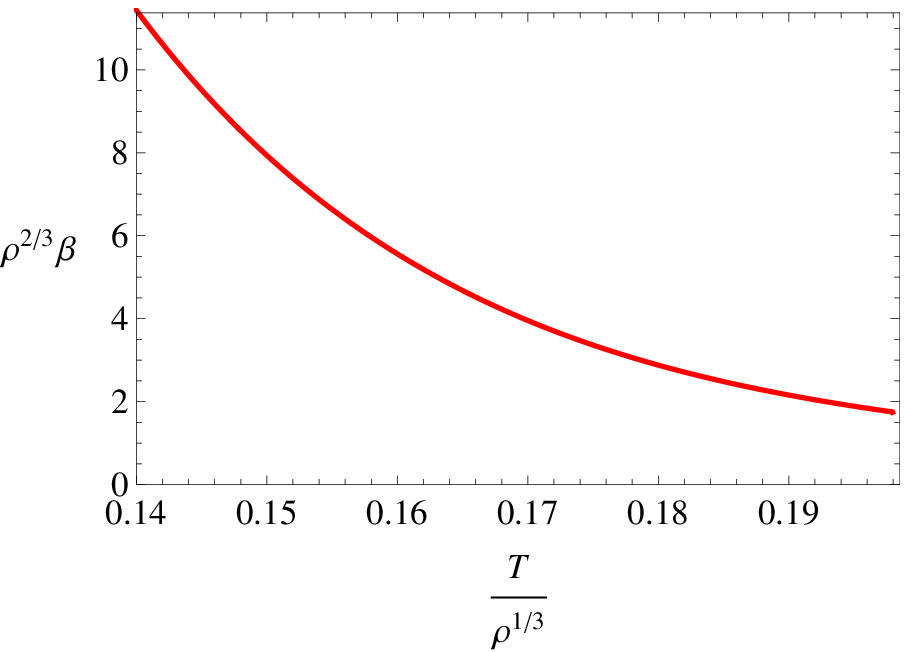}}%
\end{picture}
  \caption{$z=1$}
  \label{beta3z1}
\end{subfigure}%
\begin{subfigure}{.5\textwidth}
  \centering
	\begin{picture}(250,150)
\put(10,0){\includegraphics*[width=1\linewidth]{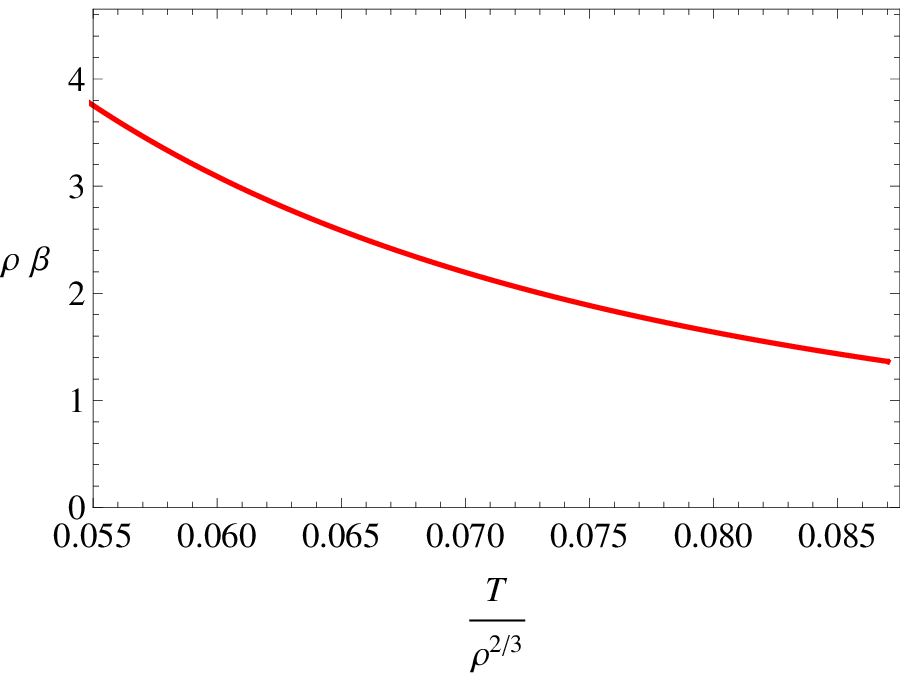}}%
\end{picture}
  \caption{$z=2$}
  \label{beta3z2}
\end{subfigure}
\caption{\textbf{Case I}.}
\label{beta3z}
\centering
\begin{subfigure}{.5\textwidth}
  \centering
 \begin{picture}(250,150)
\put(0,0){\includegraphics*[width=1\linewidth]{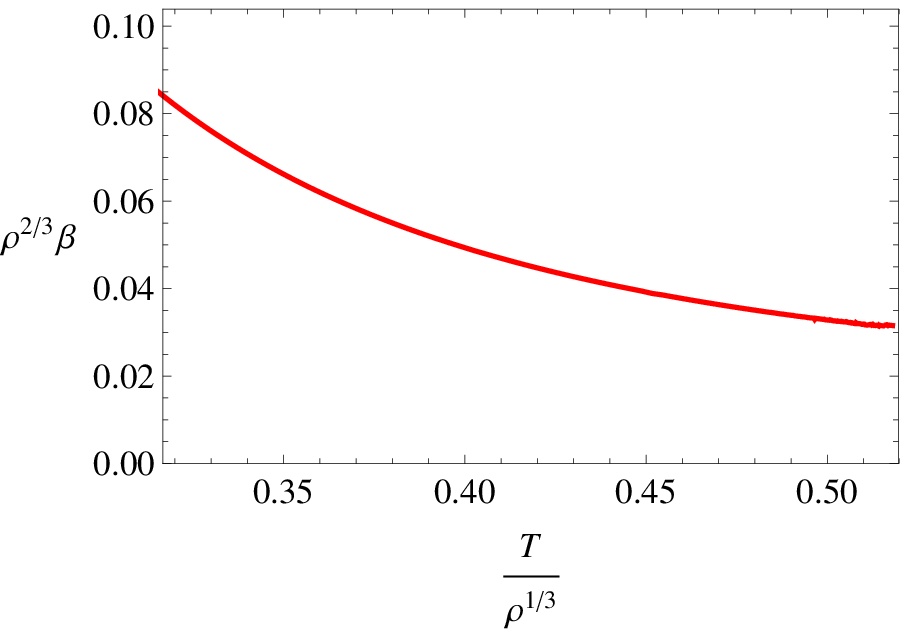}}%
\end{picture}
  \caption{$z=1$}
  \label{beta1z1}
\end{subfigure}%
\begin{subfigure}{.5\textwidth}
  \centering
	\begin{picture}(250,150)
\put(10,0){\includegraphics*[width=1\linewidth]{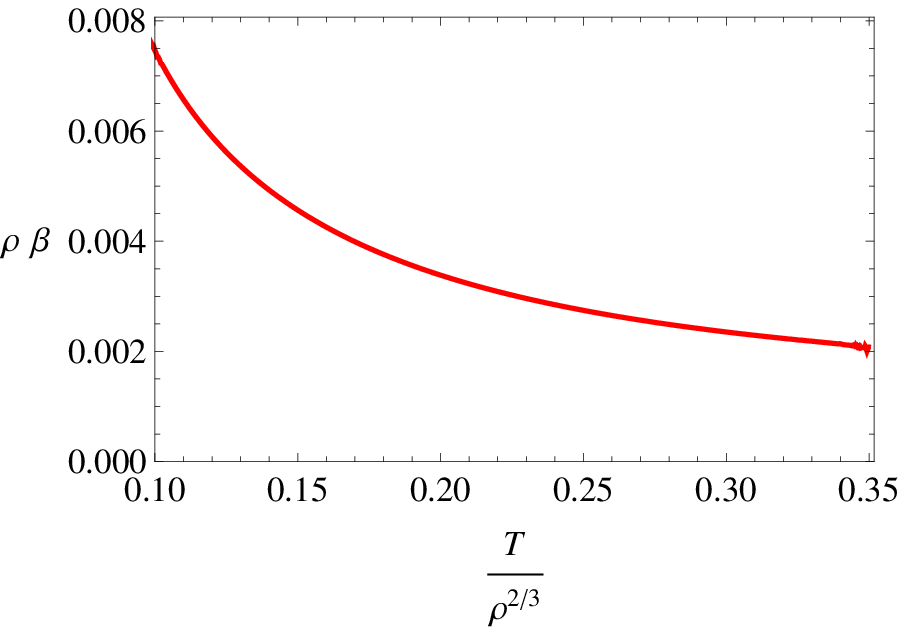}}%
\end{picture}
  \caption{$z=2$}
  \label{beta1z2}
\end{subfigure}
\caption{\textbf{Case II}.}
\label{beta1z}
\end{figure}

 The remaining phenomenological parameter $\beta$ can be computed through the Ginzburg-Landau theory relation
\begin{equation}
\label{psiinfty}
\left|\Psi_{\infty}\right|^{2}= \frac{\left|\alpha\right|}{ \beta}\,.
\end{equation}
where $\left|\Psi_{\infty}\right|$ is the value of the condensate deep inside the superconductor, where external fields and gradients are negligible. Since we are in the limit of small field perturbations, we indeed find ourselves in that approximation. Substituting (\ref{GLid}) and (\ref{alpha}) in (\ref{psiinfty}) we obtain the following expression
\begin{equation}
\label{beta}
\beta=\frac{1}{4 N_{z}}\frac{1}{\mathcal{O}_{\Delta}^{2}\xi_{0}^{2}}\,.
\end{equation}
In figures \ref{beta3z}-\ref{beta1z} we show the behavior of $\beta$ as a function of temperature, for our different condensation cases. We observe that, near-$T_{c}$, $\beta$ behaves in agreement with Ginzburg-Landau theory, having a definite value at $T=T_{c}$. We also observe that this value decreases as the value of $z$ raises. 

%\bigskip
%\bigskip
%\bigskip

\begin{figure}
\centering
\textbf{Value of the Ginzburg-Landau parameter $\kappa$ as a function of temperature, for different cases.}
\begin{subfigure}{.5\textwidth}
  \centering
 \begin{picture}(250,150)
\put(0,0){\includegraphics*[width=1\linewidth]{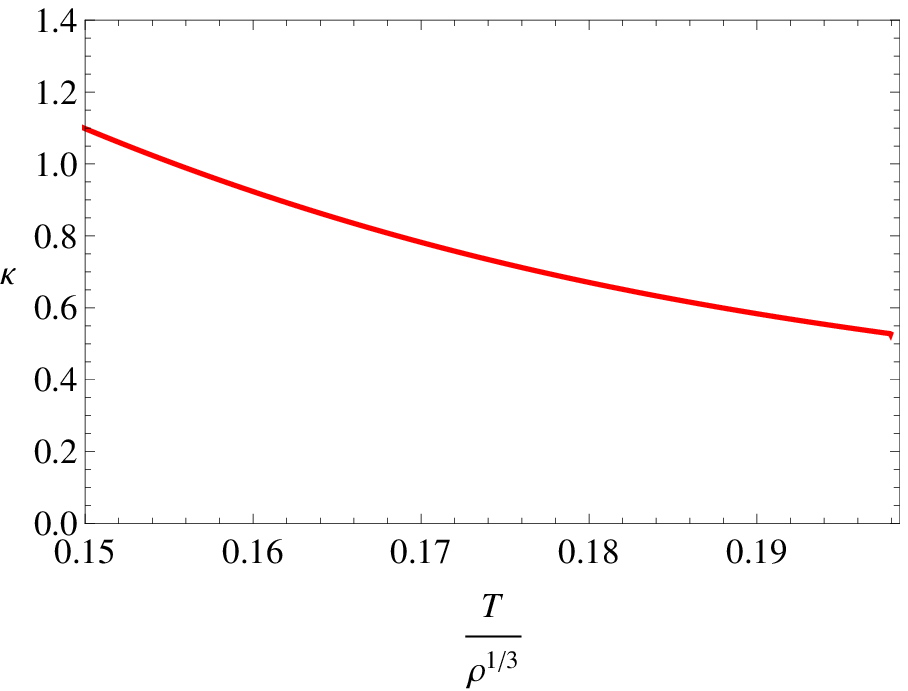}}%
\end{picture}
  \caption{$z=1$}
  \label{kappa3z1}
\end{subfigure}%
\begin{subfigure}{.5\textwidth}
  \centering
	\begin{picture}(250,150)
\put(10,0){\includegraphics*[width=1\linewidth]{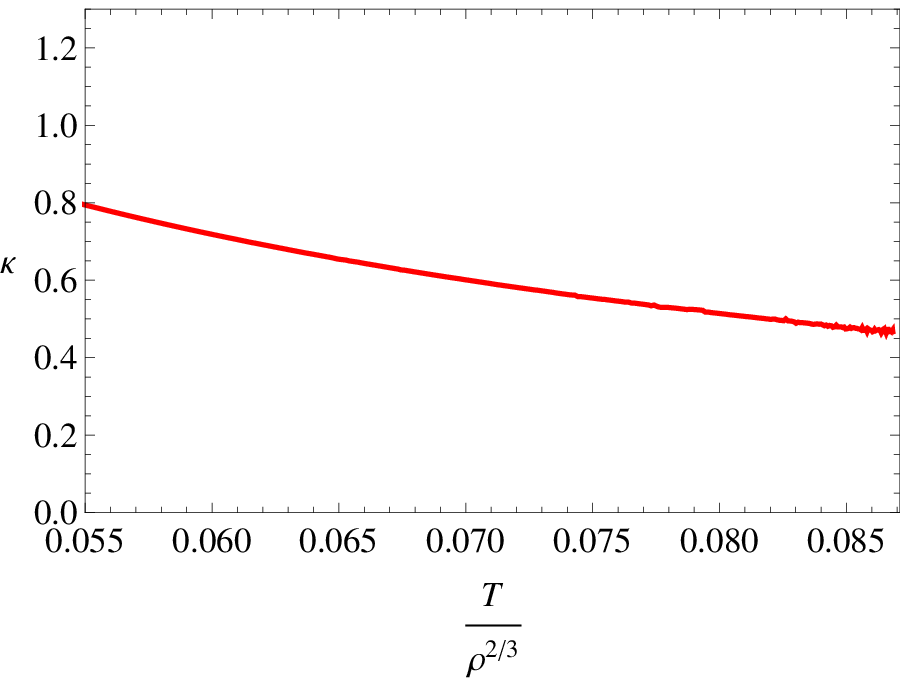}}%
\end{picture}
  \caption{$z=2$}
  \label{kappa3z2}
\end{subfigure}
\caption{\textbf{Case I}.}
\label{kappa3z}
\centering
\begin{subfigure}{.5\textwidth}
  \centering
 \begin{picture}(250,150)
\put(0,0){\includegraphics*[width=1\linewidth]{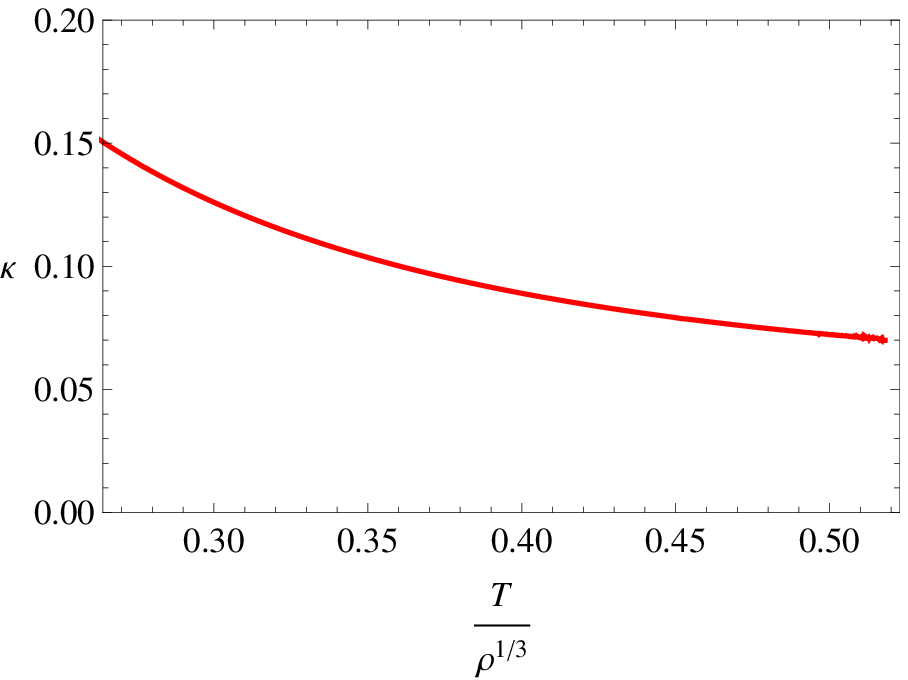}}%
\end{picture}
  \caption{$z=1$}
  \label{kappa1z1}
\end{subfigure}%
\begin{subfigure}{.5\textwidth}
  \centering
	\begin{picture}(250,150)
\put(10,0){\includegraphics*[width=1\linewidth]{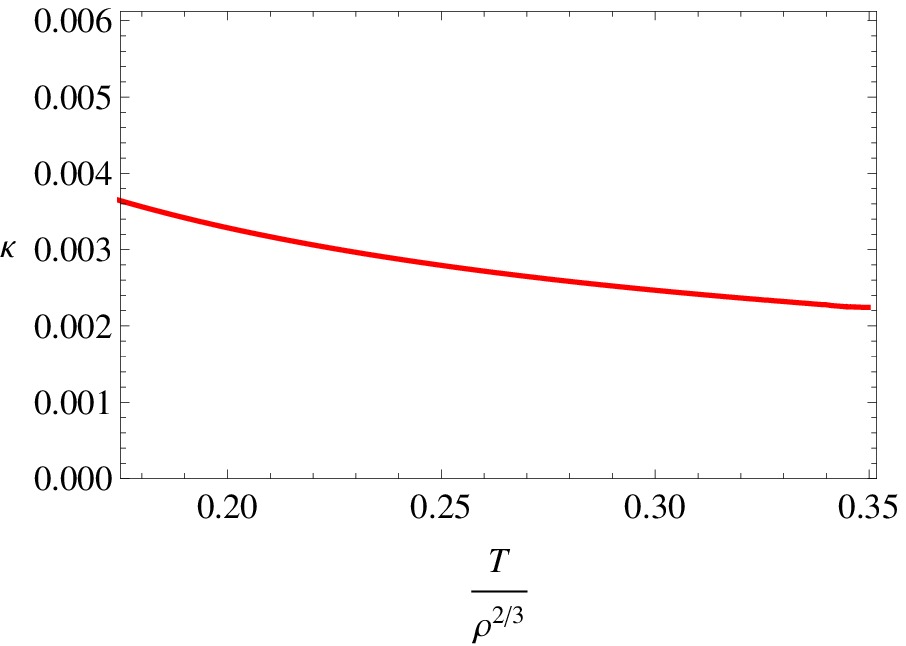}}%
\end{picture}
  \caption{$z=2$}
  \label{kappa1z2}
\end{subfigure}
\caption{\textbf{Case II}.}
\label{kappa1z}
\end{figure}

Finally, having calculated the characteristic lengths of the system $\xi_{0}$ and $\lambda$, we can compute the Ginzburg-Landau parameter $\kappa$, defined as
\begin{equation}
\kappa = \frac{\lambda}{\xi}\,,
\end{equation}
where $\xi$ is the Ginzburg-Landau coherence length, which is related to our correlation length $\xi_{0}$ as $\xi^{2}=2\xi_{0}^{2}$. (See \cite{Dector:2013dia}). In figures \ref{kappa3z}-\ref{kappa1z} we show how the Ginzburg-Landau parameter $\kappa$ behaves as a function of temperature, for our different cases. We notice that all plots have a definite value at $T=T_{c}$. We will take this to be the value of $\kappa$ of our holographic superconductor for each case considered. The value of $\kappa$ for different cases is shown in Table \ref{tabkappa}. We note that all values of $\kappa$ are lower than $1/\sqrt{2}\sim 0.707$ for all cases of $z$ considered, which means that our system behaves always as a Type I superconductor. Also, we notice that the value of $\kappa$ is always lower for $z=2$, which means that in holographic superconductors with higher dynamical critical exponent, vortex formation is more strongly unfavored energetically and have a stronger Type I behavior. 

\begin{table}[t]
\begin{center}
\caption{\label{tabkappa} Value of the Ginzburg-Landau parameter $\kappa$, for different cases.}
\begin{tabular}{ccc}
  \toprule
	\toprule
	$\kappa$& $z=1$ & $z=2$\\
	\toprule
  \textbf{Case I} & 0.527 & 0.467\\ 
  \textbf{Case II} & 0.070 & 0.002\\
  \toprule
	\toprule
\end{tabular}
\end{center}
\end{table}

%%%%%%%%%%%%%%%%%%%%%%%%%%%%%%%%%%%%%%%%%%%%%%%%%

\section{Constant Magnetic Field}

We will now study the effect of an external magnetic field to the superconducting phase of our models. As done before, we begin with in the general dimensional case and then focus on $D=5$. We follow the procedure developed by \cite{Maeda:2009vf} and will proceed in a perturbative fashion by proposing a series expansion for the component fields
\begin{eqnarray}
\label{psiexp}
\Psi(\vec{x},u)&=&\epsilon^{1/2}\Psi^{(1)}(\vec{x},u)+\epsilon^{3/2}\Psi^{(2)}(\vec{x},u)+\cdots\\
\label{phiexp}
A_{\mu}(\vec{x},u)&=&A_{\mu}^{(0)}(\vec{x},u)+\epsilon A_{\mu}^{(1)}(\vec{x},u)+\cdots
\end{eqnarray}
where $\vec{x}=(x,y)$, and the expansion parameter is given by
\begin{equation}
\epsilon=\frac{B_{c}-B}{B_{c}}\,,\hspace{25pt} \epsilon \ll 1\,,
\end{equation}
were $B_{c}$ is the value of the magnetic field that breaks the superconducting phase (\textit{critical magnetic field}). Since this expansion is done near the value $B=B_{c}$, this means that we find ourselves near the point where the condensate vanishes. We substitute expansions (\ref{psiexp})-(\ref{phiexp}) in the general equations on motion (\ref{psieom1})-(\ref{phieom1}). The zero order equation for the gauge field is
\begin{equation}
\frac{1}{\sqrt{g}}\partial_{\mu}\left(\sqrt{g}F_{(0)}^{\mu \nu}\right)=0\,,
\end{equation}
and has solutions
\begin{eqnarray}
A_{t}^{(0)}(u)&=&\mu - \rho\, \frac{u^{d-z}}{r_{h}^{d-z}}\,,\\
A_{y}^{(0)}(x)&=&B_{c}\, x\,,
\end{eqnarray}
and the rest of spatial components equal to zero: $A_{i}^{(0)}=0$, $i\neq y$. Since the solution for $A_{t}^{(0)}$ is equal to solution (\ref{phi0sol}), we set the notation $A_{t}^{(0)}=\phi$, for simplicity. Meanwhile, the general scalar field equation is
\begin{equation}
\label{psieomB}
u^{d+1-z}\partial_{u}\left(\frac{f}{u^{z+d-1}}\partial_{u}\Psi^{(1)}\right)-\left(\frac{m^{2}}{u^{2z}}-\frac{\phi^{2}}{r_{h}^{2z}f}\right)\Psi^{(1)}=-\frac{1}{r_{h}^{2}u^{2z-2}}\delta^{IJ}D_{I}D_{J}\Psi^{(1)}\,.
\end{equation}
where $I,J=x,y$. Eq. (\ref{psieomB}) is clearly separable. We follow the standard treatment and propose
\begin{equation}
\label{sol1}
\Psi^{(1)}(\vec{x},u)=e^{ipy}\varphi^{(p)}(x,u)\,,
\end{equation}
so on the right hand side of (\ref{psieomB}) we have
\begin{equation}
\delta^{IJ}D_{I}D_{J}\Psi^{(1)}=\left(\partial_{x}^{2}+\left(\partial_{y}-i B_{c}\, x\right)^{2}\right)\Psi^{(1)}=e^{ipy}\left(\partial_{x}^{2}-\left(p- B_{c}\, x\right)^{2}\right)\varphi^{(p)}\,,
\end{equation}
and we get the following equation
\begin{equation}
\label{eq1}
u^{d+1-z}\partial_{u}\left(\frac{f}{u^{z+d-1}}\partial_{u}\varphi^{(p)}\right)-\left(\frac{m^{2}}{u^{2z}}-\frac{\phi^{2}}{r_{h}^{2z}f}\right)\varphi^{(p)}=\frac{1}{r_{h}^{2}u^{2z-2}}\left(-\partial_{x}^{2}+\left(p- B_{c}\, x\right)^{2}\right)\varphi^{(p)}\,.
\end{equation}
Now, we make the separation
\begin{equation}
\label{sol2}
\varphi^{(p)}_{n}(x,u)=\rho_{n}(u)\gamma^{(p)}_{n}(x)\,,
\end{equation}
and define the variable
\begin{equation}
X=\sqrt{2  B_{c}}\left(x-\frac{p}{ B_{c}}\right)\,,
\end{equation}
so that the operator on the right hand side of (\ref{eq1}) becomes
\begin{equation}
\left[-\partial_{x}^{2}+\left(p- B_{c} x\right)^{2}\right]=(2B_{c})\left[-\partial_{X}^{2}+\frac{1}{4}X^{2}\right]\,,
\end{equation}
and acting on $\gamma_{n}^{(p)}$ we have the eigenvalue equation
\begin{equation}
\left(-\partial_{X}^{2}+\frac{1}{4}X^{2}\right)\gamma^{(p)}_{n}=\frac{\lambda_{n}}{2}\gamma^{(p)}_{n}\,,
\end{equation}
that has as a solution the eigenfunctions
\begin{equation}
\label{sol3}
\gamma_{n}^{(p)}(x)=e^{-X^{2}/4}H_{n}(X)\,,
\end{equation}
with eigenvalues
\begin{equation}
\lambda_{n}=2n+1\,,\hspace{25pt} n=0,1\ldots
\end{equation}
We choose the $n=0$ mode, which corresponds to the most stable solution \cite{Hartnoll:2008kx,Maeda:2009vf,Albash:2008eh}. As described originally in \cite{Maeda:2009vf}, the more general solution to the scalar field is given by linear superposition of the solution obtained above, with different values of $p$. (We adopt the authors notation in the following). Going back to (\ref{sol1}), (\ref{sol2}) and (\ref{sol3}), we write our solution explicitly as
\begin{equation}
\label{sol4}
\Psi^{(1)}\left(u,\vec{x}\right)=\rho_{0}(u)\sum_{l=-\infty}^{\infty} C_{l}e^{i p_{l} y}\gamma_{0}\left(x;p_{l}\right)\,,
\end{equation}   
where
\begin{equation}
\gamma_{0}\left(x;p_{l}\right)=\text{exp}\,\left\{-\frac{B_{c}}{2}\left(x-\frac{p_{l}}{B_{c}}\right)^{2}\right\}\,,
\end{equation}
and where we define
\begin{equation}
C_{l}=\text{exp}\,\left(-i\frac{\pi a_{2}}{a_{1}^{2}}l^{2}\right)\,,\hspace{20pt} p_{l}=\frac{2\pi \sqrt{B_{c}}}{a_{1}}l\,,
\end{equation}
and $a_{1}$, $a_{2}$ are real parameters. Solution (\ref{sol4}) can be rewritten as
\begin{equation}
\Psi^{(1)}\left(u,\vec{x}\right)=\frac{1}{L}\rho_{0}(u)e^{-\frac{B_{c} x^{2}}{2}}\vartheta_{3}(\upsilon,\tau)\,,
\end{equation}
where $\vartheta_{3}(\upsilon,\tau)$ is the \textit{elliptic theta function}, defined as
\begin{equation}
\vartheta_{3}(\upsilon,\tau)=\sum_{l=-\infty}^{\infty}e^{i \pi \tau l^{2}}e^{2 i \pi \upsilon l}\,,
\end{equation}
and where the variables $\upsilon$ and $\tau$ are defined as
\begin{equation}
\upsilon \equiv \frac{\sqrt{B_{c}}}{a_{1}}\left(-i x + y\right)\,,\hspace{20pt} \tau \equiv \frac{1}{a_{1}^{2}}\left(2i\pi -a_{2} \right)\,.
\end{equation}

\begin{figure}
\centering
\textbf{Value of the critical magnetic field $B_{c}$ as a function of temperature, for different cases.}
\begin{subfigure}{.5\textwidth}
  \centering
 \begin{picture}(250,150)
\put(0,0){\includegraphics*[width=1\linewidth]{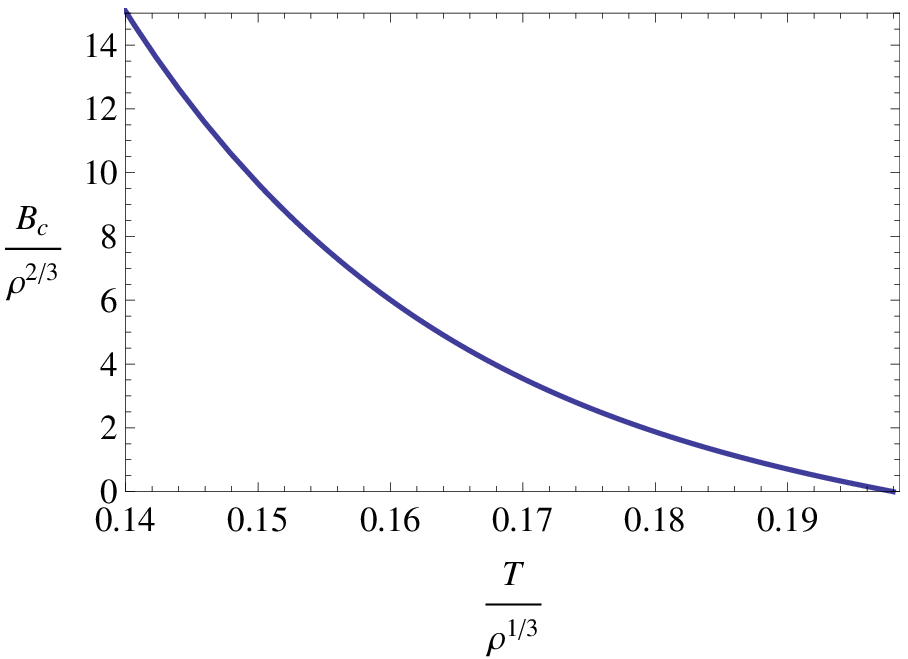}}%
\end{picture}
  \caption{$z=1$}
  \label{B3z1}
\end{subfigure}%
\begin{subfigure}{.5\textwidth}
  \centering
	\begin{picture}(250,150)
\put(10,0){\includegraphics*[width=1\linewidth]{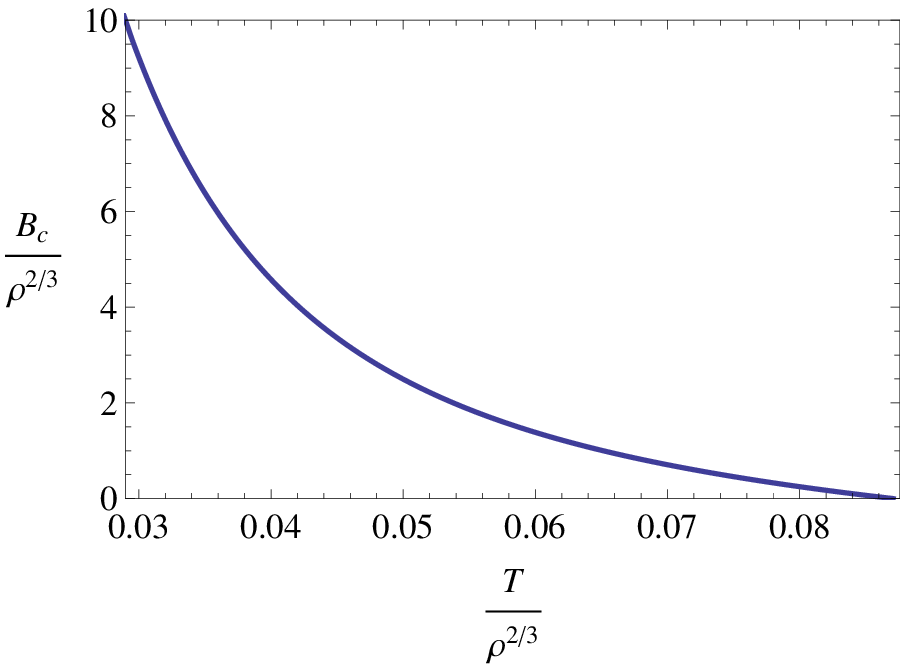}}%
\end{picture}
  \caption{$z=2$}
  \label{B3z2}
\end{subfigure}
\caption{\textbf{Case I}.}
\label{B3z}
\begin{subfigure}{.5\textwidth}
  \centering
 \begin{picture}(250,150)
\put(0,0){\includegraphics*[width=1\linewidth]{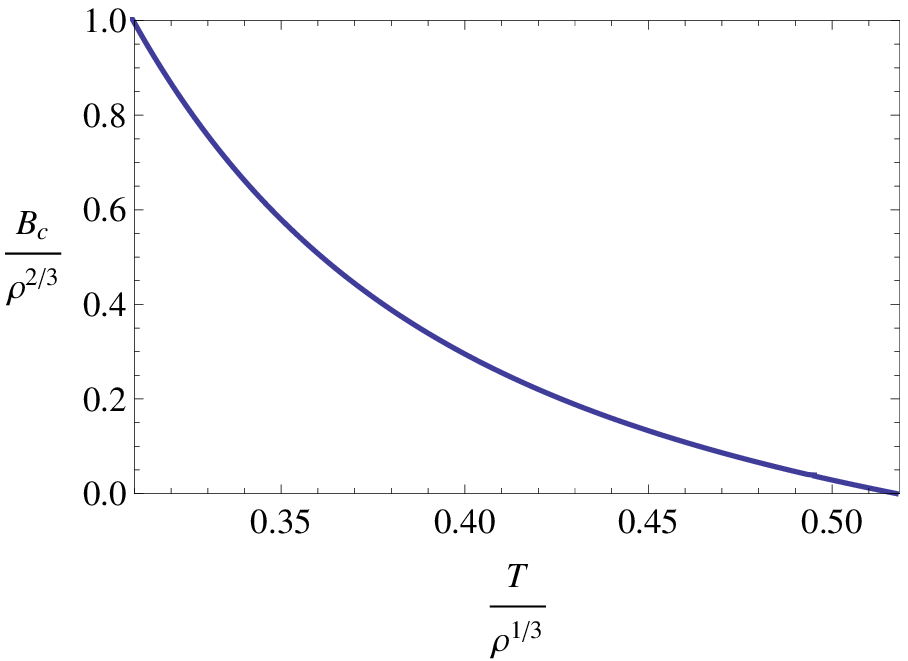}}%
\end{picture}
  \caption{$z=1$}
  \label{B1z1}
\end{subfigure}%
\begin{subfigure}{.5\textwidth}
  \centering
	\begin{picture}(250,150)
\put(10,0){\includegraphics*[width=1\linewidth]{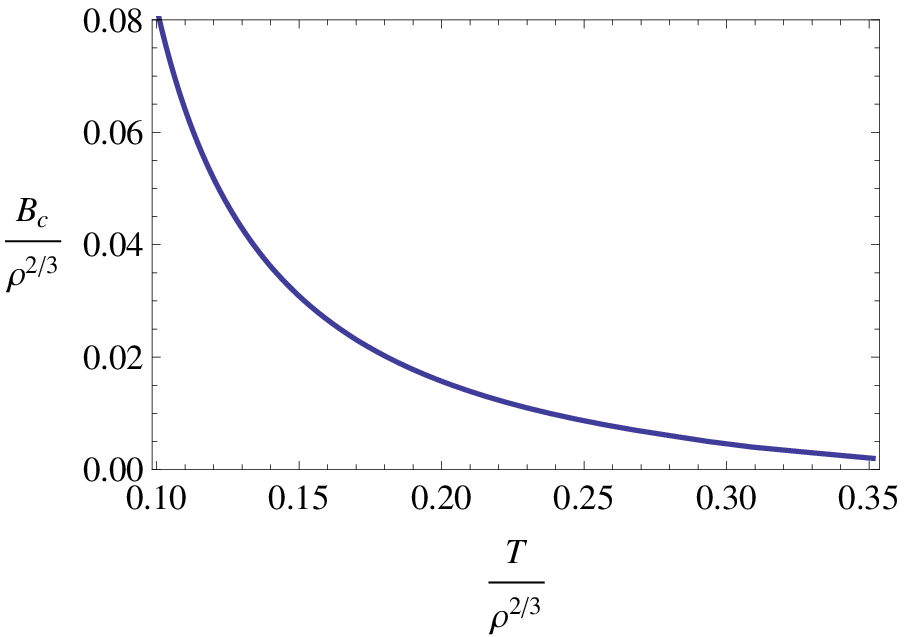}}%
\end{picture}
  \caption{$z=2$}
  \label{B1z2}
\end{subfigure}
\caption{\textbf{Case II}. }
\label{B1z}
\end{figure}

Owing to the elliptical theta function $\vartheta_{3}$, the scalar field solution $\Psi^{(1)}$ has the following pseudo-periodicity in the $x-y$ plane
\begin{eqnarray}
\Psi^{(1)}\left(u,x,y\right)&=&\Psi^{(1)}(u,x,y+a_{1})\,,\\
\Psi^{(1)}\left(u,x+\frac{2\pi}{\sqrt{B_{c}} a_{1}},y+\frac{a_{2}}{\sqrt{B_{c}}a_{1}}\right)&=&\text{exp}\,\left[\frac{2 \pi i}{a_{1}}\left(\sqrt{B_{c}}y+\frac{a_{2}}{2 a_{1}}\right)\right]\Psi^{(1)}\left(u,x,y\right)\,.\nonumber\\
\end{eqnarray}
In addition to this, the $\vartheta_{3}$ function has zeros located periodically at
\begin{equation}
\vec{V}=\left(m+\frac{1}{2}\right)\vec{v}_{1}+\left(n+\frac{1}{2}\right)\vec{v}_{2}\,,
\end{equation}
where the $\vec{v}_{i}$ vectors are given by
\begin{equation}
\vec{v}_{1}=\frac{a_{1}}{\sqrt{B_{c}}}\frac{\partial}{\partial y}\,,\hspace{20pt} \vec{v}_{2}=\frac{2\pi}{\sqrt{B_{c}}a_{1}}\frac{\partial}{\partial x}+\frac{a_{2}}{\sqrt{B_{c}}a_{2}}\frac{\partial}{\partial y}\,.
\end{equation}
Thus, the $\Psi^{(1)}$ solution has a lattice profile in the $(x-y)$ plane, spanned by the vectors $\vec{v}_{i}$. We note that, in our given approximation, we will get a 2-dimensional plane, orthogonal to the remaining $(d-2)$-dimensional boundary space, where the vortices live. We should note that the presence of the vortex solutions given above does not contradict the fact that our system was found in the previous section to be Type I\footnote{ A dynamical approach to vortex solutions in $D=4$ can be found in \cite{Dias:2013bwa}, where it was concluded that, for some values of the system's parameters, the dual superconducting system was Type I}. Indeed, the computation of $\kappa$ presented above comes from an energetic analysis, conducted directly from the dual system's Ginzburg-Landau action. (See \cite{Dector:2013dia}). This shows that, according to Ginzburg-Landau theory, the formation of the above vortex solutions costs more energy to the system than the energy needed for the system staying in a superconducting state. (See, e.g. \cite{tinkham}).

Returning to the scalar field equation (\ref{eq1}) and substituting the results given above, we obtain the following equation for the radial function $\rho$
\begin{equation}
\label{equB}
u^{d+1-z}\partial_{u}\left(\frac{f}{u^{z+d-1}}\partial_{u}\rho(u)\right)-\left(\frac{m^{2}}{u^{2z}}-\frac{\phi^{2}}{r_{h}^{2z}f}+\frac{ B_{c}}{r_{h}^{2}u^{2z-2}}\right)\rho(u)=0\,,
\end{equation}
which can be written as
\begin{equation}
\label{rhoeq}
\rho''+\left(\frac{f'}{f}-\frac{d+z-1}{u}\right)\rho'-\frac{1}{u^{2}f}\left(m^{2}-\frac{u^{2z}\phi^{2}}{r_{h}^{2z}f}+\frac{u^{2}}{r_{h}^{2}}B_{c}\right)\rho=0\,.
\end{equation}
This equation of course has the same behavior at $u \rightarrow 0$ as (\ref{assympt}) 
\begin{equation}
\rho \sim C_{-} u^{\Delta_{-}}+C_{+} u^{\Delta_{+}}\,,
\end{equation}
with $\Delta_{\pm}$ given by (\ref{Delta}). We set the same boundary conditions at $u\rightarrow 0$ as for the field $\psi$ in (\ref{fieldsanzats}). By applying the shooting method to Eq. (\ref{equB}) we find the value of the critical magnetic field that breaks the superconducting phase of the system. In figures \ref{B3z}-\ref{B1z} we show the value of the critical magnetic field $B_{c}$ as a function of temperature, for our different cases. We see that near-$T_{c}$ the critical magnetic field $B_{c}$ behaves as
\begin{equation} 
B_{c}\sim \left(1-T/T_{c}\right)\,,
\end{equation}
which is in agreement with mean field theory, for all values of $z$. We also note by comparing Eqs. (\ref{rhoeq}) and (\ref{etaeq}), that the procedure to obtain the near-$T_{c}$ values of the square of the wave number $k$ and the critical field $B_{c}$ is the same. This in turn confirms the relation between the correlation length and the critical magnetic field put forward in \cite{Lala:2014jca}
\begin{equation}
B_{c}\approx \frac{1}{\xi_{0}^{2}}\,,\hspace{20pt}(T\approx T_{c})\,.
\end{equation}

\section{Conclusions}

In this paper we have constructed a $D=5$ minimal model of holographic superconductivity in the probe limit, with a Lifshitz black hole background. Within this framework, we have studied different cases of condensation, varying within each of them the dynamical critical exponent in order to gain insight on how the system is affected by $z$ with respect to its usual isotropic behavior. We have added small scalar and gauge field fluctuations to the original component fields. These fluctuations allow us to compute holographically  the penetration and coherence length of the superconducting system. We saw that both characteristic lengths have the standard near-$T_{c}$ functional dependency on temperature for all condensate cases and all values of $z$. However, the dynamical critical  exponent $z$ does affect the value of the characteristic lengths, as it becomes evident in the change of the value of their ratio as given by the Ginzburg-Landau parameter $\kappa$.  We also saw that it is possible to construct a consistent Ginzburg-Landau phenomenological interpretation of the dual theory with Lifshitz scaling. We computed through holographic techniques the Ginzburg-Landau Lagrangian parameters $\alpha$, $\beta$ and, as with the characteristic lengths, concluded that they have the standard near-$T_{c}$ functional dependency on temperature for all condensate cases and all values of $z$. However, the presence of $z$ does have a non-trivial effect on this phenomenological parameters, diminishing the value of their numerical coefficients as $z$ raises.

We have also computed with holographic techniques the Ginzburg-Landau parameter $\kappa$ of the system. For all case of condensation and all values of $z$, we saw that $\kappa<1/\sqrt{2}$. This means that for all cases the dual system will behave as a Type I superconductor. Moreover, we also observed that, for each case of condensation considered, the value of $\kappa$ became lower for higher values of $z$. This means that in systems with higher anisotropy, vortex formation is more strongly unfavored energetically and exhibit a stronger Type I behavior.

Finally, we computed the critical magnetic field $B_{c}$ needed to break the superconducting phase of the system, following the perturbative procedure first developed in \cite{Maeda:2009vf}. We observed that the critical field near-$T_{c}$ functional dependence on temperature is the one predicted by Ginzburg-Landau theory. However, we also note that the value of the critical magnetic field is smaller for higher values of $z$. Additionally, within this perturbative approach, we have confirmed holographically the conjecture posed in \cite{Lala:2014jca} that the critical magnetic field is inversely proportional to the square of the correlation length, in accordance to Ginzburg-Landau theory.

All of the above results were obtained from a minimal model of superconductivity following \cite{Hartnoll:2008kx}. It would be interesting to see how these results would be affected by the choice of other models, such as, for instance, d-wave holographic superconductors \cite{Gubser:2008wv}, models with higher corrections to the scalar field potential such as the ones that appear in top-down approaches \cite{Gubser:2009qm, Aprile:2011uq, Aprile:2012ai} or less conventional models such as ones with Chern-Simons terms, higher-derivative couplings or in the context of New Massive Gravity \cite{Banerjee:2013maa, Kuang:2013oqa, Abdalla:2013zra}.
%%%%%%%%%%%%%%%%%%%%%%%%%%%%%%%%%%%%%%%%%%%%%%%%%%%%%%%%%%%%%%%%%%%%%%%%%%%%%%%%%%%%%%%%%%%%%%%%%%%%%%%%%%%%%%%%%%%%%%%%%%%%%%

\section*{Acknowledgments}

It is a pleasure to thank J. G. Russo for his guidance and encouragement during the research of this project. I also wish to thank F. Aprile for very valuable suggestions. This research was funded by CONACyT grant No.306769.

%%%%%%%%%%%%%%%%%%%%%%%%%%%%%%%%%%%%%%%%%%%%%%%%%%%%%%%%%%%%%%%%%%%%%%%%%%%%%%%%%%%%%%%%%%%%%%%%%%%%%%%%%%%%%%%%%%%%%%%%%%%%

%%%%%%%%%%%%%%%%%%%%%%%%%%%%%%%%%%%%%%%%%%%%

%%%%%%%%%%%%%%%%%%%%%%%%%%%%%%%%%%%%%%%%%%%%
\end{document}